\documentclass[
    prd, twocolumn, superscriptaddress, nofootinbib, amsmath, amssymb,
    aps, floatfix
]{revtex4-2}
\usepackage[utf8]{inputenc}
\usepackage[
    colorlinks=true,
    linkcolor=blue,
    urlcolor=blue,
    citecolor=blue
]{hyperref}

\usepackage[utf8]{inputenc}
\usepackage{setspace,flushend}
\usepackage[dvips]{graphicx}
\usepackage[dvipsnames]{xcolor}
\usepackage{tikz}
\usepackage{soul}

\usepackage{amsmath,amssymb,mathrsfs,layout,textcomp}

\newcommand{\refcite}[1]{Ref.~\cite{#1}}
\newcommand{\refscite}[1]{Refs.~\cite{#1}}

\usepackage{bm}
\newcommand{\bb}[1]{\bm{\mathrm{#1}}}
\newcommand{\bt}[1]{\bm{\mathrm{\mathsf{#1}}}}

\usepackage{siunitx}
\DeclareSIUnit{\year}{yr}
\newcommand{\eV}{\electronvolt}
\newcommand{\meV}{\milli\eV}
\newcommand{\keV}{\kilo\eV}
\newcommand{\MeV}{\mega\eV}

\usepackage{mhchem}
\newcommand{\URuSi}{\ce{URu_2Si_2}}

\newcommand{\du}{\mathrm d}
\newcommand{\dd}{\,\du}

\newcommand{\fermi}{\mathrm{F}}
\newcommand{\lon}{\mathrm{L}}
\newcommand{\plas}{\mathrm{p}}
\newcommand{\dm}{\chi}
\newcommand{\med}{\phi}

\newcommand{\tens}[1]{\overset{\leftrightarrow}{\bt{#1}}}
\newcommand{\iso}{\mathrm{iso}}

\usepackage{amsfonts}
\usepackage{mathtools}
\usepackage{graphicx}
\usepackage{mathrsfs}
\usepackage{slashed}
\usepackage{esint}
\usepackage{mathtools}
\definecolor{citeblue}{RGB}{45,52,151}
\usepackage{hyperref}
\hypersetup{colorlinks = true, allcolors = citeblue}

\usepackage[capitalise]{cleveref}
\crefname{sec}{Section}{Sections}

\newcommand{\bpm}{\begin{pmatrix}}
    \newcommand{\lp}{\left(}
    \newcommand{\re}{\text{Re}\,}

    \newcommand{\im}{\text{Im}\,}
    \newcommand{\lb}{\left[}
    
    \newcommand{\rb}{\right]}

    \newcommand{\rp}{\right)}
    \newcommand{\e}{\epsilon}
    \newcommand{\epm}{\end{pmatrix}}

\DeclarePairedDelimiter\bra{\langle}{\rvert}
\DeclarePairedDelimiter\ket{\lvert}{\rangle}
\DeclarePairedDelimiterX\braket[2]{\langle}{\rangle}{#1 \delimsize\vert #2}
\newcommand{\loss}{\R_{\mathrm{EELS}}(\bb q,\omega)}

\newcommand{\R}{\mathcal W}

\begin{document}

\title{Directional detection of dark matter with anisotropic response functions}

\author{Christian Boyd}
\affiliation{Department of Physics, University of Illinois Urbana-Champaign, Urbana, IL 61801, USA}

\author{Yonit Hochberg}
\affiliation{Racah Institute of Physics, Hebrew University of Jerusalem, Jerusalem 91904, Israel}

\author{Yonatan Kahn}
\affiliation{Department of Physics, University of Illinois Urbana-Champaign, Urbana, IL 61801, USA}
\affiliation{Illinois Center for Advanced Studies of the Universe, University of Illinois Urbana-Champaign, Urbana, IL 61801, USA}

\author{Eric David Kramer}
\affiliation{Jerusalem College of Technology, Jerusalem 93721, Israel}
\affiliation{Racah Institute of Physics, Hebrew University of Jerusalem, Jerusalem 91904, Israel}

\author{Noah Kurinsky}
\affiliation{SLAC National Accelerator Laboratory, Menlo Park, CA 94025, USA}
\affiliation{Kavli Institute for Particle Astrophysics and Cosmology, Stanford Univeristy, Stanford, CA 94305, USA}

\author{Benjamin V. Lehmann}
\affiliation{Center for Theoretical Physics, Massachusetts Institute of Technology, Cambridge, MA 02139, USA}

\author{To Chin Yu}
\affiliation{SLAC National Accelerator Laboratory, Menlo Park, CA 94025, USA}
\affiliation{Kavli Institute for Particle Astrophysics and Cosmology, Stanford Univeristy, Stanford, CA 94305, USA}

\date{\today}
\begin{abstract}
    Direct detection for sub-GeV dark matter is developing rapidly, with many novel experimental ideas and theoretical methods emerging. In this work, we extend the dielectric formalism for dark matter scattering to incorporate anisotropic material responses, enabling directionally-sensitive experiments with a broad class of target materials. Using a simple model of an anisotropic electron gas, we demonstrate the importance of many-body effects such as the plasmon, and show that even when the dark matter kinetic energies are much smaller than the plasmon energy, the tail of an anisotropic plasmon can still produce a sizeable daily modulation. We highlight the relevant experimental techniques required to establish the target response, as well as the challenges in extracting a response function which is truly free of modeling uncertainties.
\end{abstract}

\maketitle

\section{Introduction}
Over the past several decades, laboratory searches have played a crucial role in constraining dark matter (DM) models at the GeV scale and above. These experiments are based on the detection of very rare interactions between DM and Standard Model (SM) particles \cite{Alexander:2016aln,Battaglieri:2017aum}, meaning that background reduction is crucial \cite{Akerib:2022ort,Baxter:2022dkm}. Despite considerable efforts by experimental collaborations, it is generally difficult to ensure that all backgrounds have been eliminated, and as detection thresholds are lowered, unexpected backgrounds are inevitably uncovered.

Nonetheless, it is possible to preserve sensitivity to a DM signal by leveraging the properties of the DM distribution in the laboratory. Since the velocity distribution of DM particles is isotropic in the galactic frame, Earth's motion induces a DM ``wind'' with a direction that changes over the course of a sidereal day. Detectors with sensitivity to the direction of the incoming DM particles can thus identify a rare DM signal even in the presence of a background. For GeV-scale and heavier DM, experiments typically achieve this directional sensitivity by observing the direction of a recoiling final-state nucleus. Several proposed experiments would be capable of detecting such a modulation~\cite{Avignone:2008cw,Drukier:2012hj,Mayet:2016zxu,NEWSdm:2017efa,Beaufort:2021uyg,Ebadi:2022axg}.

However, at lower masses, DM detection with or without directional sensitivity remains a significant challenge. Below the GeV scale, the minuscule kinetic energy and momentum deposited by a DM particle are no longer large compared to those of single excitations in the detector system, so the separation of scales assumed in traditional experiments no longer applies. In this regime, the condensed matter physics of the detector system becomes significant to the DM scattering rate \cite{Essig:2022dfa,Kahn:2022kae,Mitridate:2022tnv}. In particular, a new generation of experiments is making fast progress in probing DM--electron interactions down to the MeV scale and below \cite{LUX:2018akb,XENON:2019gfn,SENSEI:2020dpa,SuperCDMS:2020ymb,SuperCDMS:2020aus,EDELWEISS:2020fxc,Hochberg:2021yud}, where the response of the electron system has enormous implications for the DM scattering rate. On the other hand, directional sensitivity becomes somewhat easier at low masses because anisotropic targets can have excitation rates which depend on the incoming DM direction, such that observation of the direction of a final-state recoil is not necessary. There are a number of proposals which could achieve directional sensitivity for sub-GeV DM \cite{Hochberg:2016ntt,Cavoto:2017otc,Hochberg:2017wce,Budnik:2017sbu,Kadribasic:2017obi,Griffin:2018bjn,Heikinheimo:2019lwg,Coskuner:2019odd,Geilhufe:2019ndy,Coskuner:2021qxo,Sassi:2021umf,Blanco:2021hlm,Hochberg:2021ymx,Blanco:2022pkt}.

For spin-independent interactions, the relationship between the DM scattering rate and the electronic density response in dielectric targets was recently established by \refcite{Hochberg:2021pkt} (see also \refcite{Knapen:2021run}), providing a new set of heuristics for the connection between material properties and the sensitivity of DM searches. Here we extend the formalism of \refcite{Hochberg:2021pkt} to anisotropic systems, laying the groundwork for the development of anisotropic low-threshold detectors from arbitrary target materials.

In this work, we directly relate the event rate of DM--electron interactions and associated daily modulation to causal many-body response functions of condensed matter systems. We demonstrate this with a toy model of electronic anisotropy, the anisotropic electron gas, for which we directly compute the response function and extract the modulation amplitude. Beyond simple models, the utility of our framework is that the material properties governing DM--electron scattering rates, including anisotropic response, can be measured through electron energy loss spectroscopy (hereafter EELS). Our formalism thus provides a pathway to  determining daily modulation rates in arbitrary condensed matter systems, particularly those for which simple models or computational techniques such as density functional theory~(DFT) are unreliable or infeasible, so long as EELS data can be obtained in the momentum and energy regime of interest.

We focus throughout on the dynamic structure factor $S(\bb q,\omega)$, where $\bb q$ and $\omega$ denote the deposited momentum and energy, respectively, and we work in the scattering regime where $|\bb q| \gg \omega$. The dynamic structure factor directly determines the DM--electron scattering rate~\cite{Trickle:2019nya}. \refcite{Hochberg:2021pkt} writes the dynamic structure factor in terms of the dielectric function $\epsilon(\bb q, \omega)$, which is the longitudinal component of the dielectric tensor $\tens \e(\bb q, \omega)$. This allowed for a rapid transfer of heuristics from the theory of dielectric screening to the computation of the DM scattering rate. Here, in order to clarify the origin of anisotropic response, we frame the discussion in terms of the causal electronic density response, a scalar function defined by
\begin{equation}
    \label{eq:density-response-definition}
    \chi(\bb r,\bb r';t) \equiv -i\Theta(t)
    \bra0\left[
        \hat n(\bb r,t),\,\hat n(\bb r',t)
    \right]
    \ket0
    ,
\end{equation}
where $\Theta(t)$ is the Heaviside function, $\hat n(\bb r)$ is the electron density operator, and $|0 \rangle$ is the many-body ground state of the target. Generically, $S(\bb q,\omega)$ is related to the imaginary part of the density response function by the fluctuation--dissipation theorem, which we will consider explicitly in the zero temperature limit, as appropriate for cryogenic low-threshold detectors.

The density response approach is just as compatible with experimental calibration as the formulation based on the dielectric tensor. Indeed, the dielectric tensor is defined in terms of the in-medium electric field, which is not directly measurable at the momenta and energies relevant for DM scattering, and requires further manipulation in order to yield the scattering rate. We stress that within the approximations of this work, the formulations in terms of $S(\bb q, \omega)$ and $\epsilon(\bb q, \omega)$ are equivalent. The relationships between the various quantities used to describe the material response are illustrated in \cref{fig:scattering-functions}. We additionally emphasize that when EELS data is not yet accessible, it is sufficient for DFT practitioners to compute the \emph{scalar} dynamic structure factor rather than the dielectric \emph{tensor}, even for the purposes of assessing anisotropic response.

This paper is organized as follows.  In \cref{sec:scattering}, we review the basic scattering formalism and establish the connection between DM--electron scattering and electron--electron scattering through the density response function.  In \cref{sec:aeg}, we compute the response function for a toy model, an electron gas with an anisotropic effective electron mass. In \cref{sec:daily-modulation}, we compute the associated daily modulation for the anisotropic electron gas for various DM masses and mediator masses, illustrating the interplay between the DM energy scales and the condensed matter energy scales, characterized by the  plasmon energy and the Fermi velocity. In \cref{sec:calibration}, we discuss the feasibility of directly measuring the anisotropic response function with EELS, and the limitations one can expect to encounter in practice. We conclude in \cref{sec:conclusions}.

\section{Dark matter scattering and the density response function}
\label{sec:scattering}

\begin{figure*}[!t]
    \centering
    \includegraphics[width=\textwidth]{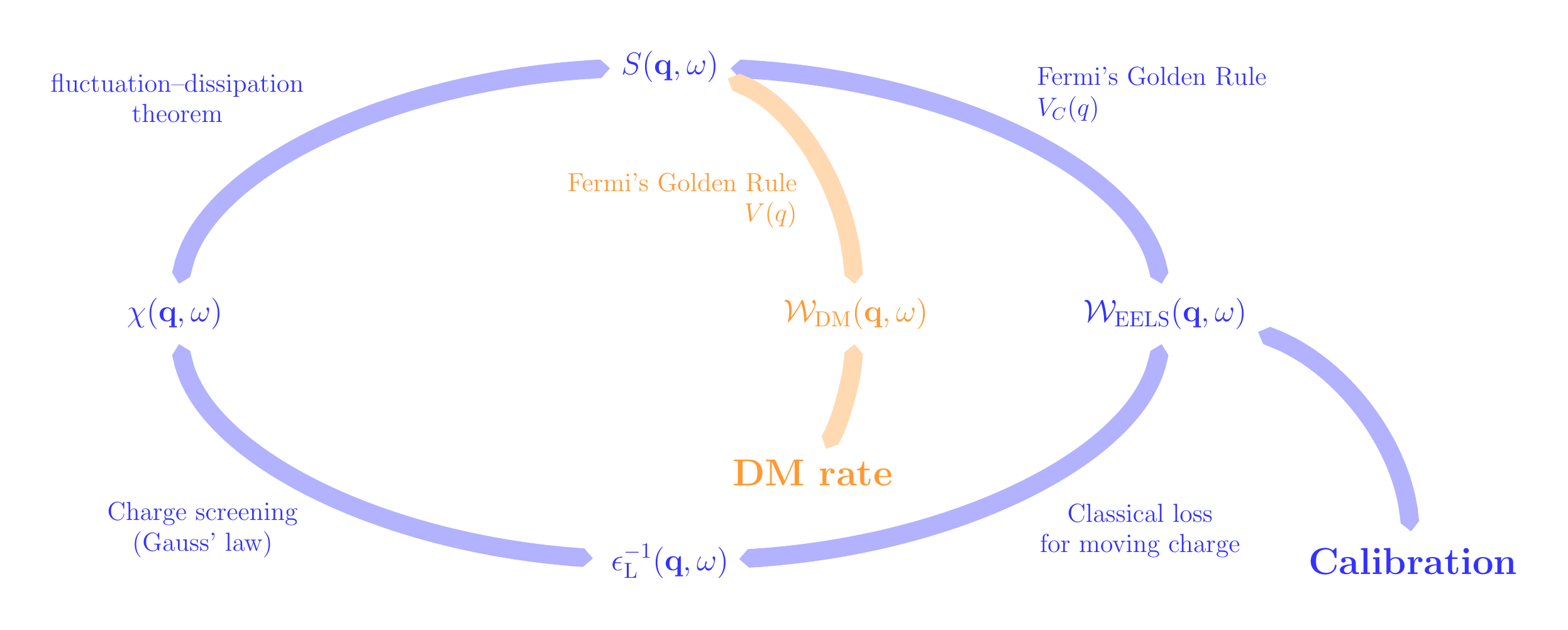}
    \caption{
 Schematic depiction of the relationships between the various quantities governing the target response and the DM scattering rate.  Fermi's Golden Rule provides a momentum- and energy-resolved scattering rate in terms of the target dynamic structure factor $S(\bb q,\omega)$.  The density response function $\chi(\bb q,\omega)$ of a target is related to $S(\bb q,\omega)$ by the fluctuation--dissipation theorem, \cref{eq:structure-factor-density-response}.  The inverse longitudinal dielectric function $\e^{-1}_\lon(\bb q,\omega)$ is determined through Gauss's law by the charge density response function, which is proportional to $\chi(\bb q,\omega)$ for an electronic system.  As $\e^{-1}_\lon(\bb q,\omega)$ and $\chi(\bb q,\omega)$ are causal response functions, they are entirely determined by their (dissipative) imaginary parts through Kramers--Kronig relations. Importantly, these quantities can be measured with calibration experiments, and then related directly to the DM scattering rate in the material.
    }
    \label{fig:scattering-functions}
\end{figure*}

We begin by reviewing the basic framework for scattering in solid-state targets, the connection between DM--electron and electron--electron scattering, and the properties of the local DM distribution. Our emphasis here is to phrase these well-known aspects in terms of the key theoretical quantity, the density response function, the imaginary part of which yields the dynamic structure factor $S(\bb q,\omega)$. (For earlier methods, see \refscite{Trickle:2019nya,Kahn2022} and references therein.) We use natural units throughout, with $\hbar=\e_0=c=1$.

We focus exclusively on a spin-independent density--density interactions between DM and electrons, with momentum-space potential of the form
\begin{equation}
    \label{eq:potential-definition}
    V(q) = \frac{g_e g_\chi}{m_\phi^2+q^2}
    ,
\end{equation}
depending only on the magnitude of the momentum transfer $q = |\bb q|$, where $\phi$ denotes a scalar (vector) mediator with Yukawa (gauge) coupling $g_e$ to electrons and $g_\chi$ to DM. Since galactic DM is non-relativistic, the scattering rate for a stream of DM incident on a solid state target can be computed using the non-relativistic Hamiltonian coupling DM to the target electron density,
\begin{equation}
    \label{eq:interaction-hamiltonian}
    H_{\mathrm{int}} = \int \du^3\bb r \, \hat n(\bb r) \, \Phi(\bb r)
    ,
\end{equation}
where $\Phi(\bb r)$ is the potential sourced by the incident DM position-space configuration, and $\hat n(\bb r)$ is the electron density operator. The condensed matter physics of the target enters the DM--electron scattering rate as computed by Fermi's Golden Rule through the electron dynamic structure factor,
\begin{equation}
    \label{eq:structure-factor}
    S(\bb q,\omega) \equiv \frac{2\pi}{\Omega}\sum_{f}
    \bigl|\bra{f}\,\hat n(-\bb q)\,\ket{0}\bigr|^2
    \delta\bigl(\omega - \left[E_\fermi-E_0\right]\bigr)
    \,.
\end{equation}
Here $\Omega$ is the material volume, the sum runs over all excited states $\ket f$, $E_f-E_0$ is the energy difference between $\ket f$ and $\ket 0$, $\bb q$ is the momentum transfer to the target, and $\omega$ is the energy loss of the scattered DM. For additional details, see \cref{sec:material-response} or e.g. \refcite{Sturm1993}\footnote{We match the conventions of \refcite{Kahn2022}, including a factor of $2\pi$ in $S(\bb q,\omega)$ associated with Fermi's Golden rule. This is different from the conventional usage of the fluctuation--dissipation theorem in solid-state physics (e.g. as in \refcite{Sturm1993}).}. In writing \cref{eq:structure-factor}, we are explicitly taking the solid-state target to be in its ground state $\ket0$ at zero temperature. In this case, the DM particle can only lose energy to the target, so that $\omega>0$.

The structure factor depends only on the properties of the target, regardless of the nature of the incoming SM or DM particle. The fluctuation--dissipation theorem imposes a relationship between $S(\bb q, \omega)$ and the electronic density response function $\chi(\bb q,\omega)$, which takes the following form at zero temperature:
\begin{equation}
    \label{eq:structure-factor-density-response}
    S(\bb q,\omega) = -2\,\im\chi(\bb q,\omega)
    .
\end{equation}
We emphasize that $S(\bb q,\omega)$ and $\chi(\bb q,\omega)$ are scalar functions, since they encode the effects of spin-independent scattering mediated via a density-density interaction. Anisotropy in the response arises only from the dependence of these scalars on the direction of $\bb q$, contrary to what one might expect when formulating the calculation in terms of the tensorial dielectric function.\footnote{Note that $S$ and $\chi$ are mathematically equivalent: the analyticity properties of $S$ and $\chi$ imply that the imaginary part of $\chi$ is sufficient to reconstruct the entire function through the Kramers--Kronig relations.}

Notably, the coupling in \cref{eq:interaction-hamiltonian} is not unique to DM--electron scattering. Electron scattering from a solid-state target takes place via precisely the same coupling \cite{nozieres1959electron}. As a result, electron--electron scattering also probes the structure factor of \cref{eq:structure-factor} encountered in DM--electron scattering.  In electron--electron scattering, however, the perturbing potential is due to the long-ranged Coulomb interaction,
\begin{equation}
    \label{eq:coulomb-potential}
    V_C(q)= \frac{e^2}{q^2}
    \;,
\end{equation}
between the probe and material electrons.  This correspondence between DM--electron and electron--electron scattering, as first detailed in \refscite{Hochberg:2021pkt,Knapen:2021run}, allows us to make contact with the vast condensed matter literature on the scattering of electrons with solid-state targets when modeling the DM interaction.

In EELS experiments, fast electrons are shot through thin material foils and analyzed according to their angular deflection and energy loss.  As a result, EELS is an energy- and momentum-resolved probe of charge density excitations in solids \cite{Egerton2011,Fink2014}, analogous to deep inelastic scattering in nuclear physics for probing the structure of nucleons.\footnote{In this work we neglect both phonon excitation and the DM--ion coupling; if the latter coupling exists, phonons are expected to contribute at the 10--\SI{100}{\meV} scale in the total charge density response probed by EELS, and in gapped materials would dominate below the gap if the gap is well above the typical phonon energy.}  The energy loss spectrum in EELS is determined by the electron energy loss function \cite{Pines1959,Egerton2011,Fink2014} which, through the relations in \cref{fig:scattering-functions}, can be written as
\begin{equation}
    \label{eq:loss-response-dielectric}
    \R_{\mathrm{EELS}}(\bb q,\omega)=-V_C(q)\,\im\chi(\bb q,\omega)= -\im \e_\lon^{-1}(\bb q,\omega)
    \,\,,
    \end{equation}
where $\e_\lon \equiv \bb{\hat q} \cdot \tens\epsilon \cdot\bb{\hat q}$ is the longitudinal dielectric function of the solid state target. In what follows, we will assume that the target possesses translation invariance, as is the case for our anisotropic electron gas model in \cref{sec:aeg}, so that $\chi(\bb r,\bb r'; t)=\chi(\bb r-\bb r',t)$. However, the wavevector-dependent quantities within the EELS loss function of \cref{eq:loss-response-dielectric} and \cref{fig:scattering-functions} generalize to macroscopic crystalline solids, as these relations do not rely on translation symmetry; see e.g.\ the construction in \cref{app: scattering} for $S(\bb q,\omega)$ and $\chi(\bb q,\omega)$.

While the connection between density response and dielectric screening is a foundational accomplishment of many-body theory, the relation between the electron energy loss function and the longitudinal inverse dielectric function in \cref{eq:loss-response-dielectric} will play an auxiliary, rather than essential, role in our analysis.  In \cref{fig:scattering-functions}, we outline the physical relationships between the various quantities in our discussion of electron scattering which carry the same information: the longitudinal inverse dielectric function $\e^{-1}_\lon(\bb q,\omega)$, the density response function $\chi(\bb q,\omega)$, the dynamic structure factor $S(\bb q,\omega)$, and the EELS loss function $\loss$. As any one of these quantities unambiguously determines the others, our modeling efforts will focus on the object which is readily calculated by standard methods: the density response function $\chi(\bb q,\omega)$, as presented in \cref{sec:aeg}.  The string of equalities in \cref{fig:scattering-functions} extends just as well to DM-electron scattering, which we illustrate by introducing the DM energy loss function,
\begin{equation}
    \label{eq:dm-loss-function}
    \R_{\mathrm{DM}}(\bb q,\omega) = -V(q)\im\chi(\bb q,\omega)
    ,
\end{equation}
constructed by analogy to the EELS loss function in \cref{eq:loss-response-dielectric} through the DM-electron interaction of \cref{eq:potential-definition}.  The choice of DM model dictates the interaction $V(q)$, at which point the target response---including all anisotropic and many-body effects---is determined entirely by the scalar function $\chi(\bb q,\omega)$ through the relations in \cref{fig:scattering-functions}.  Of crucial practical importance is that the target contribution to DM-electron scattering is identical to the response one can extract from calibration experiments using EELS. 

Given $S(\bb q,\omega)$, the DM--electron scattering rate can be readily computed. For a DM particle moving with velocity $\bb v_\chi$ through a solid state target, the scattering rate per unit target mass is given by Fermi's Golden Rule as~\cite{Trickle:2019nya}
\begin{equation}
    \label{eq:dm-scattering-rate}
    \Gamma(\bb v_\dm) = \frac{\pi\bar\sigma_e}{\mu_{e\dm}^2}
    \int\frac{\du^3\bb q \dd\omega}
    {(2\pi)^3}
    \mathcal F(q)^2
    S(\bb q, \omega)\,
    \delta\bigl(\omega-\omega_{\bb q}(\bb v_\dm)\bigr)
    ,
\end{equation}
where $\mu_{e\dm}$ is the reduced mass of the electron--DM system; $\bar\sigma_e \equiv \frac1\pi\mu_{e\dm}^2g_e^2g_\dm^2/(m_\med^2 + q_0^2)^2$ is a reference cross section, with $q_0 =\alpha m_e$ a benchmark momentum transfer; $\mathcal F(q) \equiv V(q) / V(q_0) = (m_\med^2 + q_0^2)/(m_\med^2 + q^2)$ parameterizes the momentum dependence of the DM--electron potential; and $\omega_{\bb q}$ is the deposited energy at fixed $\bb q$, given explicitly by
\begin{equation}
    \label{eq:dm-energy-loss}
    \omega_{\bb q}(\bb v_\dm) = \bb q\cdot\bb v_\dm - \bb q^2/2m_\dm
    .
\end{equation}
Further details can be found e.g. in the review of \refcite{Kahn2022} and in \cref{sec:SHM}. As previously emphasized, we will focus on the zero-temperature limit in which $S(\bb q, \omega)$ only has support for $\omega > 0$.

Naturally, the DM velocity is not a constant in the lab frame, but takes the form of a velocity distribution $f_{\mathrm{lab}}(\bb v_\dm,t)$ that changes with time as the Earth rotates relative to our galactic DM halo. Integrating \cref{eq:dm-scattering-rate} over this distribution gives the total rate as a function of time:

\begin{equation}
\label{eq:event-rate}
    R(t) = \frac{1}{\rho_T}\frac{\rho_\dm}{m_\dm}
        \frac{\pi\bar\sigma_e}{\mu_{e\dm}^2}
        \int\frac{\du^3\bb q\dd\omega}{(2\pi)^3}g_0(\bb q, \omega,t)
        \mathcal F(q)^2
            S(\bb q, \omega)
    .
\end{equation}
where, following \refcite{Trickle:2019nya}, we parameterize the DM halo integral with
\begin{equation}
\label{eq:g0}
    g_0(\bb q, \omega,t) =
    \int\du^3\bb v_\dm\, f_{\mathrm{lab}}(\bb v_\dm,t) \,
    \delta\lp\omega - \omega_{\bb q}(\bb v_\chi)\rp
    .
\end{equation}
Use of \cref{eq:event-rate} allows one to directly extract the daily modulation of the rate in any target material---including the full many-body response of the system---given $S(\bb q,\omega)$ obtained by either experimental or computational means.

\section{Response of an anisotropic electron gas}
\label{sec:aeg}

As a concrete model of an electronic system with anisotropic density response, we consider an electron gas with the anisotropic dispersion
\begin{equation}
    \label{eq:aeg-dispersion}
    E_{\bb q} = 
    \frac{q^2_x}{2m_x}+\frac{q^2_y}{2m_y} + \frac{q^2_z}{2m_z}
    ,
\end{equation}
where $\bb q$ corresponds to the 3D wavevector of an electron quasiparticle and the unequal $m_{x,y,z}$ are effective masses describing band anisotropy along three orthogonal spatial directions. As noted in \refscite{Ruvalds1977,Campos1989,dasSarma2021}, the anisotropic dispersion in \cref{eq:aeg-dispersion} can be related to an isotropic system by use of a scale transformation,
\begin{equation}
    \label{eq:anisotropic-rescaling}
    q_i\to Q_i(\bb q)=q_i\sqrt{\frac{M}{m_i}}
\end{equation}
applied along each of the primary axes of the material, i.e., $i\in\{x,y,z\}$. Here $M$ is a density-of-states (DOS) mass,
\begin{equation}
    \label{eq:mass-definition}
    M= \left(m_x m_y m_z\right)^{1/3}
    .
\end{equation}
In what follows, we will denote rescaled wavevectors with capital letters. For example, given a wavevector $\bb p$, the version rescaled according to \cref{eq:anisotropic-rescaling} is denoted by $\bb P$.  Using the definitions in \cref{eq:anisotropic-rescaling,eq:mass-definition}, we can write the anisotropic dispersion of \cref{eq:aeg-dispersion} as
\begin{equation}
    \label{eq:anisotropic-relation}
    E_{\bb q}
    = \frac{q^2_x}{2m_x}+\frac{q^2_y}{2m_y} + \frac{q^2_z}{2m_z}
    = \frac{Q^2}{2M}
    ,
\end{equation}
which can in turn be written as the dispersion of a related isotropic system as
\begin{equation}
    \label{eq:isotropic-dispersion}
    E_{\bb Q}^{\iso} = \frac{Q^2}{2M}
    .
\end{equation}
Notably, \cref{eq:anisotropic-rescaling} and \cref{eq:anisotropic-relation} are consistent independent of the definition of $M$.  The DOS mass in \cref{eq:mass-definition}, as used by \refcite{dasSarma2021}, is chosen such that the transformation of \cref{eq:anisotropic-rescaling} has unit determinant, i.e., such that
\begin{equation}
    \label{eq:unit-determinant}
    \dd^3\bb q = \dd^3 \bb Q.
\end{equation}

As established in \cref{sec:scattering}, the solid state target determines the DM--electron scattering through the dynamic electronic structure factor, $S(\bb q,\omega)$, which is related to the density response function, $\chi(\bb q,\omega)$, via \cref{eq:structure-factor-density-response}.  The density response function of our non-interacting, single-band electron gas model is given by 
\begin{equation}
    \label{eq:chi0-definition}
    \chi_0(\bb q,\omega) =
    2\int\frac{\dd^3 \bb p}{(2\pi)^3}
    \frac{n_\fermi(E_{\bb p+\bb q})-n_\fermi(E_{\bb p})}
         {E_{\bb p+\bb q}-E_{\bb p}-\omega-i\Gamma_\plas}
    ,
\end{equation}
where the factor of $2$ accounts for the electron spin degeneracy within the dispersion of \cref{eq:aeg-dispersion}, and $\Gamma_\plas>0$ phenomenologically sets the linewidth of resonant excitations.  In keeping with the conventions of \cref{sec:scattering}, we restrict our analysis to the zero-temperature limit in which the Fermi distribution function, $n_\fermi$ in \cref{eq:chi0-definition}, behaves as a step function,
\begin{equation}
    n_\fermi(E) =
    \begin{cases}
    1 & E<E_\fermi \\
    0 & E>E_\fermi,
    \end{cases}
\end{equation}
in terms of the Fermi energy of our system, $E_\fermi$.  We now utilize the unimodular scale transformation in \cref{eq:anisotropic-rescaling} to rewrite \cref{eq:chi0-definition} as
\begin{equation}
    \label{eq:anisotropic-to-isotropic}
    \chi_0(\bb q,\omega) =
    \chi_0^{\iso}\bigl(\bb Q(\bb q), \omega\bigr)
    ,
\end{equation}
where
\begin{equation}
    \label{eq:chi0-iso-def}
    \chi^{\iso}_{0}(\bb Q,\omega)
    =
    2\int\frac{\dd^3 \bb P}{(2\pi)^3}
    \frac{n_\fermi(E^{\iso}_{\bb P+\bb Q})-n_\fermi(E_{\bb P}^{\iso})}
         {E^{\iso}_{\bb P+\bb Q}-E^{\iso}_{\bb P}-\omega-i\Gamma_\plas}
\end{equation}
is the non-interacting density response function of an \emph{isotropic} electron gas with band mass given by \cref{eq:mass-definition}.  The relation in \cref{eq:anisotropic-to-isotropic} provides a great simplification, as the density response function of a free electron gas is a textbook result \cite{Mahan2000,Bruus2004}.\footnote{A similar scaling trick was used to compute the response of anisotropic Dirac materials in \refcite{Hochberg:2017wce}. }

\begin{figure*}[t!]
    \centering
    \includegraphics[width=\textwidth]{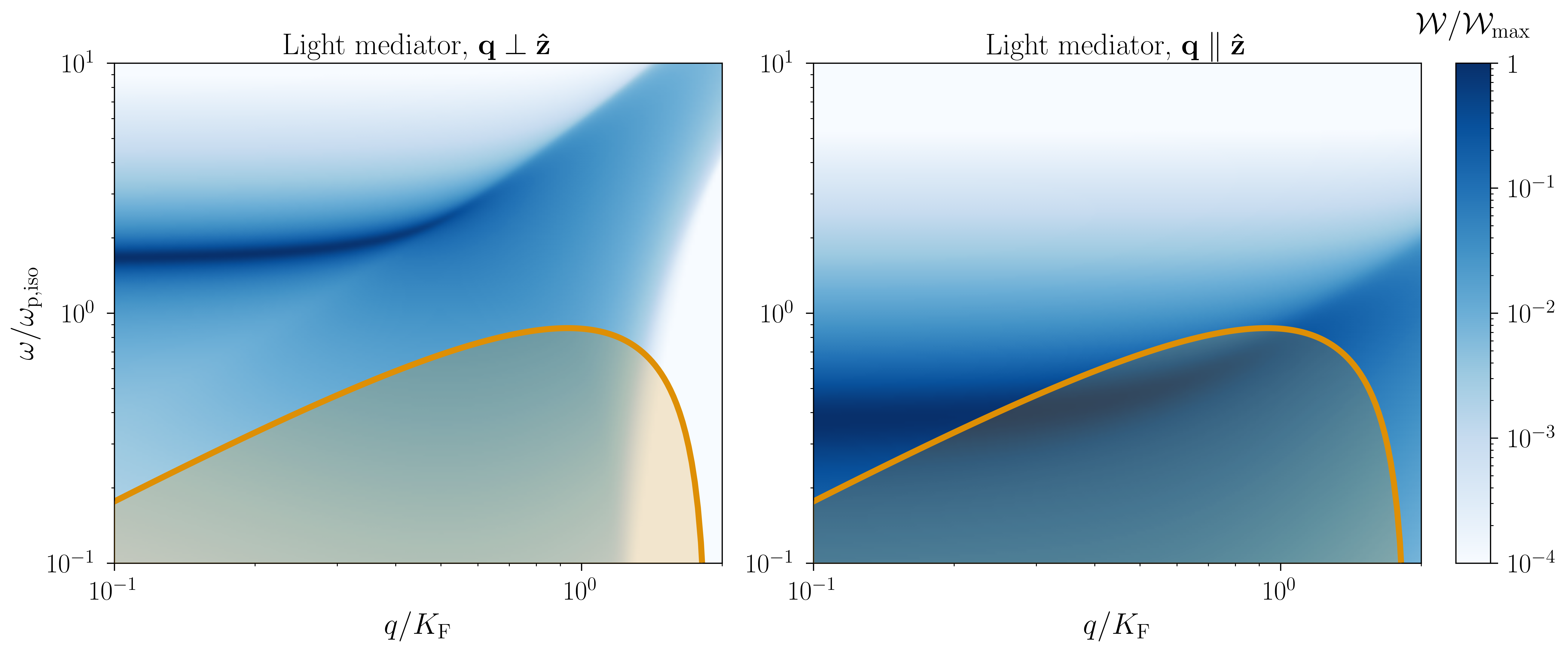}
    \caption{Density plot (shaded blue) of the DM loss function for a light mediator, $\R_{\mathrm{DM}}(\bb q,\omega)$ of \cref{eq:dm-loss-function}, determined by the RPA density response function (\cref{eq:model-RPA}) of an anisotropic electron gas with $\epsilon_\infty$ set to ensure $\omega_{\plas,\iso} = E_\fermi$, measured for momentum transfers perpendicular \textbf{(left)} and parallel \textbf{(right)} to $\hat{\mathbf{z}}$. The anisotropy parameters were chosen to be $m_x=m_y=0.37M$ and $m_z=7.37M$, such that $m_z/m_{xy}=20$ and $\lp m_xm_ym_z\rp^{1/3}=M$. The dark blue band illustrates the plasmon dispersion $\Omega(q)$, which is at higher frequencies along the light mass directions and lower frequencies along the heavy mass direction, approaching $\Omega_{xy} = 1.6 \omega_{\plas,\iso}$ and $\Omega_z = 0.37 \omega_{\plas,\iso}$ as $q \to 0$. To make contact with the physical scale set by the DM dispersion $v_0$, we show the kinematically-allowed region for DM scattering with $m_\chi = m_e/2$ and $v_\chi = \SI{800}{\kilo\meter/\second}$ in shaded orange. With respect to these scales, the electron gas parameters should be taken as $M = m_e$, $E_\fermi = \omega_{\plas,\iso} = \SI{1}{\eV}$, and $\epsilon_\infty = 6.26$, which we choose as our fiducial values for the remainder of this paper unless otherwise specified. With these parameters, only the lower plasmon branch (right) is energetically accessible.}
    \label{fig:loss-function}
\end{figure*}

Though $\chi_0(\bb q,\omega)$ in \cref{eq:chi0-definition} is the response function that corresponds to our anisotropic electron gas model of \cref{eq:aeg-dispersion}, real materials feature many-electron screening due to the long-ranged Coulomb interaction. Since our electron gas model is, at best, an effective model of the valence electrons in an anisotropic target, we model screening by the core electrons\footnote{See, e.g., the use of $\e_\text{core}$ in the free carrier models of \refcite{Dresselhaus2018}.} through an effective Coulomb interaction with background dielectric constant $\e_\infty$,
\begin{equation}
    \label{eq:coulomb-eff}
    V_{C}^{\mathrm{eff}}(q) =
    \frac{V_C(q)}{\e_\infty} = \frac{e^2}{\e_\infty q^2}
    .
\end{equation}
The inclusion of $\e_\infty$ will also provide us more control over the relevant energy scales in the problem.  The many-electron screening within our electron gas model is most simply treated self-consistently through the Random Phase Approximation (RPA) (see e.g. \refscite{Sturm1993,Pines1988,Mahan2000,Bruus2004}). Within RPA, the density response function can be written as 
\begin{equation}
    \label{eq:RPA}
    \chi_{\mathrm{RPA}}(\bb q,\omega) =
    \frac{\chi_0(\bb q,\omega)}
         {1-V_C^{\mathrm{eff}}(q)\chi_0(\bb q,\omega)}
    .
\end{equation}
Using RPA is equivalent to summing the bubble diagrams that result from a perturbative expansion for $\chi(\bb q,\omega)$ in the presence of the effective Coulomb interaction, $V_C^{\mathrm{eff}}$ of \cref{eq:coulomb-eff}.  If we combine the isotropic relation in \cref{eq:anisotropic-to-isotropic} with the RPA expression of \cref{eq:RPA}, the RPA density response function of our anisotropic electron gas can be written in terms of the isotropic response function of \cref{eq:chi0-iso-def} as
\begin{equation}
    \label{eq:model-RPA}
    \chi_{\mathrm{RPA}}(\bb q,\omega) =
    \frac{\chi_0^{\iso}(\bb Q(\bb q), \omega)}
         {1-V_C^{\mathrm{eff}}(q)\chi_0^{\iso}(\bb Q(\bb q), \omega)}
    .
\end{equation}

As \cref{eq:structure-factor-density-response} allows us to determine the dynamic structure factor $S(\bb q,\omega)$ through $\im\chi_{\mathrm{RPA}}(\bb q,\omega)$, we are in a position to discern the kinds of excitations that would be visible in a scattering experiment. Particle--hole excitations can be seen directly in $-\im\chi_0^{\iso}(\bb q,\omega)$, while many-electron excitations (i.e. plasmons) appear whenever the resonance condition,
\begin{equation}
    \label{eq:plasmon-condition}
    1-V_C^{\mathrm{eff}}(q) \re\chi_0^{\iso}(\bb Q(\bb q), \omega) = 0
    ,
\end{equation}
is met. In writing \cref{eq:plasmon-condition} as the implicit dispersion relation for plasmon excitation, we are assuming a small width, or $\Gamma_\plas/\omega\ll1$ at the frequency satisfying the pole condition.  Notably, there is a peak contribution to $\im\chi_{\mathrm{RPA}}(\bb q,\omega)$ at the plasmon resonance condition of \cref{eq:plasmon-condition} even as $\Gamma_p\to0^+$ through the Dirac identity (c.f. \cref{app: Dirac identity}). Accounting for the plasmon becomes essential at small $q$, or small $Q$ in $\chi_0^{\iso}(\bb Q(\bb q),\omega)$, where the phase space for particle-hole excitation is restricted to a thin shell about the Fermi surface and the plasmon contribution dominates.  

In the regime of plasmon propagation, we can find the long-wavelength plasmon dispersion relation directly from the asymptotic form of $\re\chi_0^{\iso}$,
\begin{equation}
    \label{eq:chi0-long-wavelength}
    \re\chi_0^{\iso}(\bb Q,\omega)
    \overset{Q_i\to0}\longrightarrow
    \frac{n_eQ^2}{M\omega^2}
    .
\end{equation}
Inserting this into \cref{eq:plasmon-condition} yields the long-wavelength plasmon dispersion as
\begin{equation}
    \label{eq:plasmon-general-dispersion}
    \omega^2_\plas(\bb q)=\frac{n_e e^2}{M\e_\infty}\frac{Q^2(\bb q)}{q^2}
    .
\end{equation}
In \cref{eq:chi0-long-wavelength,eq:plasmon-general-dispersion}, $n_e=K_{\fermi}^3/3\pi^2$ is the electron density, where $K_{\fermi}$ is the \emph{isotropic} Fermi wavevector defined through \cref{eq:isotropic-dispersion} by $E_{K_\fermi}^{\iso}=E_\fermi$, namely
\begin{equation}
\label{eq:KFdef}
K_\fermi = \sqrt{2M E_\fermi}.
\end{equation}
Conveniently, the unimodular transformation of \cref{eq:unit-determinant} between the anisotropic and isotropic systems allows us to calculate $n_e$ in terms of isotropic parameters. The scale of $\omega_\plas$ in \cref{eq:plasmon-general-dispersion} is provided by the plasma frequency of the isotropic system,
\begin{equation}
    \label{eq:isotropic-plasmon}
    \omega^2_{\plas,\iso}=\frac{n_ee^2}{M\e_\infty} =
    \frac{K_\fermi^3 e^2}{3\pi^2 M\e_\infty}
    ,
\end{equation}
which reduces to the well-known plasma frequency\footnote{The frequency $\omega_{\plas,\iso}$ in \cref{eq:isotropic-plasmon} is occasionally termed the \emph{screened} plasma frequency since it includes the background dielectric constant.  As \cref{eq:isotropic-plasmon} corresponds to the physical frequency of the plasmon excitation in our model, we suppress the ``screened'' moniker.} in the isotropic limit of equal mass parameters $m_{x,y,z}$ in \cref{eq:aeg-dispersion} \cite{Pines1952,Pines1959}.  In the anisotropic system, the scale transformation of \cref{eq:anisotropic-rescaling} applied to \cref{eq:plasmon-general-dispersion} provides the plasmon dispersion as
\begin{equation}
    \label{eq:anisotropic-plasmon-dispersion}
    \omega^2_\plas(\bb q) =
    \frac{Q^2(\bb q)}{q^2}\,\omega_{\plas,\iso}^2 =
    \frac{q_x^2\Omega^2_{x} +q_y^2\Omega^2_{y}+q_z^2\Omega^2_z}{q^2}
    ,
\end{equation}
where the frequency scale along each axis, $\Omega_{x,y,z}$, is defined in terms of the mass parameters by
\begin{equation}
    \label{eq:axis-plasma-frequency}
    \Omega^2_i = \frac{n_e e^2}{m_i\e_\infty} =
    \lp\frac{M}{m_i}\rp\omega_{\plas,\iso}^2
    .
\end{equation}

\begin{figure*}[t!]
    \centering
    \includegraphics[width=\textwidth]{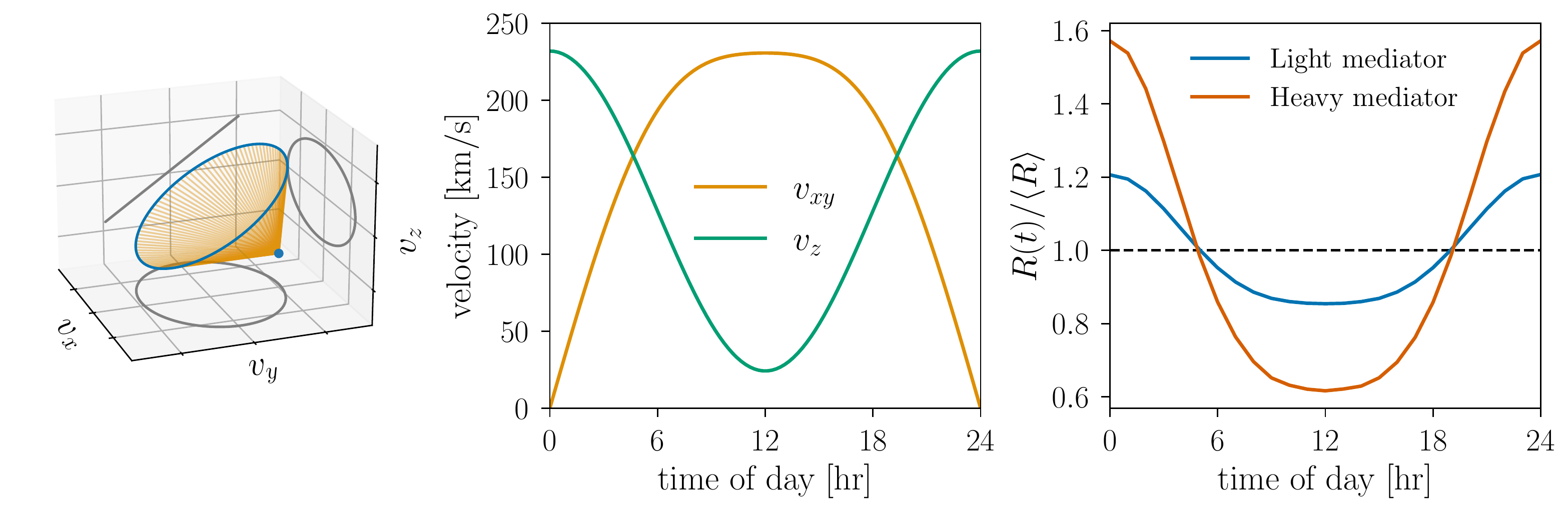}
    \caption{\textbf{Left:} 
    orientation of the lab velocity $\bb v_{\mathrm{lab}}$ of \cref{eq:v-lab} over a 24 hour period.
    \textbf{Center:} 
    as the anisotropy considered in \cref{fig:loss-function} is uniaxial, we plot for illustration purposes the daily modulation of in-plane $v_{xy}$ and out-of-plane $v_z$ components of ${\bb v}_{\mathrm{lab}}$.
    \textbf{Right:}
    daily modulation in the event rate $R(t)$ of \cref{event rate example} with a DM stream, for the same anisotropy parameters considered in \cref{fig:loss-function} including the \SI{10}{\meV} cutoff mimicking experimental resolution. The $y$-axis has been normalized to the time-averaged event rate $\langle R\rangle$.  
    }
    \label{fig:v-lab}
\end{figure*}

As an example of an appreciable but realistic mass anisotropy, we consider a uniaxial system with $m_x=m_y \equiv m_{xy} = 0.37M$ and $m_z=7.37M$, such that $m_z/m_{xy}=20$ and $\lp m_xm_ym_z\rp^{1/3}=M$. To simplify calculations, we set a single energy scale in the electronic system by using $\e_\infty$ to fix $\omega_{\plas,\iso}=E_\fermi$. \Cref{fig:loss-function} shows the DM loss function, $\R_{\mathrm{DM}}(\bb q,\omega)$ of \cref{eq:dm-loss-function}, for this system calculated using the RPA density response function of \cref{eq:model-RPA}. Notably, $\R_{\mathrm{EELS}}(\bb q,\omega)$ of \cref{eq:loss-response-dielectric} is proportional to $\R_{\mathrm{DM}}(\bb q,\omega)$ for a massless mediator. The energy and momentum scales are normalized to $\omega_{\plas,\iso}$ of \cref{eq:isotropic-plasmon} and the isotropic Fermi wavevector $K_\fermi$ of \cref{eq:KFdef}, respectively. According to our long wavelength plasmon dispersion in \cref{eq:plasmon-general-dispersion}, these mass parameters give plasma frequencies $\Omega_x = \Omega_y  \equiv \Omega_{xy} = 1.6\omega_{\plas,\iso}$ and $\Omega_z= 0.37 \omega_{\plas,\iso}$. The plasmon peak appears as a dense ridge at finite energies in both panels of \cref{fig:loss-function} as $q\to 0$. As $q$ increases, the plasmon eventually intersects the particle-hole continuum, which provides a decay channel through Landau-damping until the plasmon eventually ceases to propagate.  As noted in \refcite{dasSarma2021}, the regime of plasmon propagation is dependent upon the propagation direction, as opposed to the textbook case of an isotropic electron gas \cite{Pines1959,Mahan2000,Bruus2004}.

To make contact with the relevant DM kinematics, we can fix the physical scales of the system to some representative values: we take $E_\fermi = \SI{1}{\eV}$ and $M = m_e = \SI{511}{\keV}$, which fixes $\epsilon_\infty = 6.26$. This yields an isotropic Fermi velocity $V_\fermi \equiv \sqrt{2E_\fermi/M}=K_\fermi/M$ which is numerically equal to 590 km/s, a factor of a few larger than $v_0 = \SI{220}{\kilo\meter/\second}$ which is the physical scale of the galactic DM velocity. The $|\bb q|\to0$ plasma frequency in the $xy$-plane is $\Omega_{xy} = \SI{1.6}{\eV}$ compared to $\Omega_z = \SI{0.37}{\eV}$ along the $\hat{\bb z}$-axis. For reference, the maximum DM kinetic energy for $m_\chi= m_e/2$ is $\frac{1}{2}m_\chi v_{\mathrm{max}}^2 \simeq \SI{0.9}{\eV}$, where $v_{\mathrm{max}} \simeq \SI{800}{\kilo\meter/\second}$ is the maximum DM speed in the lab frame. For DM of this mass, the lower branch of the plasmon is energetically accessible at nonzero $q$, while the upper branch is inaccessible for all $q$. We illustrate this feature in \cref{fig:loss-function} by drawing the DM energy-loss curve $\omega_{\bb q}$ from \cref{eq:dm-energy-loss} for $m_\chi = m_e/2$ and $v_\chi = \SI{800}{\kilo\meter/\second}$ in orange. To represent the effects of finite detector resolution, we will also implement when necessary a low-energy cutoff $\omega_{\mathrm{min}} = \SI{10}{\meV}$, which is safely below any of the interesting scales in the problem (and indeed lies below the bottom of the plot in \cref{fig:loss-function}).

We close this section by noting that, while the plasmon spectrum in \cref{eq:anisotropic-plasmon-dispersion} follows from a concrete calculation within RPA, it is also quite generic to anisotropic systems of sufficient symmetry.  As detailed in~\cite{dasSarma2021}, the long wavelength plasma frequency of an anisotropic electron gas reduces to the isotropic result of \cref{eq:isotropic-plasmon} along each axis.  This reduction can be further generalized to orthorhombic Fermi surfaces~\cite{Kulik1962} and the more extreme anisotropy of layered materials~\cite{Grecu1973}.  Recently, \refcite{Lasenby2022} considered extrinsically-tuned anisotropy in the form of metal-insulator heterostructures. The resulting plasmon spectrum in these systems \cite{Mills1984,Cottam1989,Cottam1993,Cottam2004} is similar to our anisotropic plasmon dispersion in \cref{eq:anisotropic-plasmon-dispersion} if we take the uniaxial limit $m_x=m_y \equiv m_{xy}$ and consider strong anisotropy ($m_{xy}/m_z\ll 1$) in \cref{eq:aeg-dispersion}.  While the quantitative details of any particular anisotropic material will necessarily go beyond our electron gas model in \cref{eq:aeg-dispersion}, the anisotropic plasmon dispersion in \cref{eq:anisotropic-plasmon-dispersion} provides behavior that may be qualitatively sufficient at long wavelengths.  For example, a scenario in which one branch of the plasmon is kinematically accessible to light DM, as illustrated in the right panel of  \cref{fig:loss-function}, may be realized for heavy-fermion materials~\cite{Hochberg:2021pkt}.

\section{Daily Modulation}
\label{sec:daily-modulation}

\begin{figure*}[t!]
    \centering
    \includegraphics[width=\textwidth]{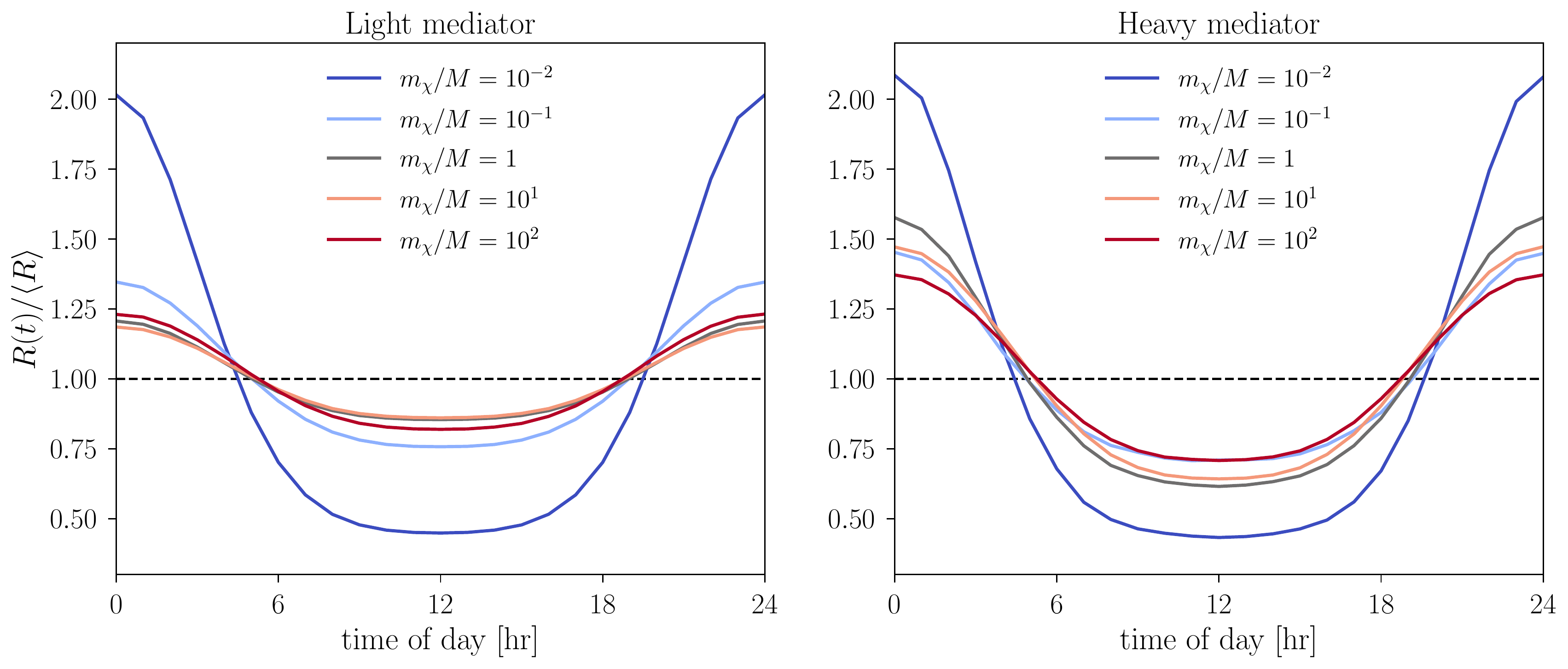}
    \caption{ 
    Daily modulation curves for the anisotropic electron gas target with DM following the SHM velocity distribution, for various values of the DM mass and a light mediator \textbf{(left)} and heavy mediator \textbf{(right)}. We take the target to have the same fiducial parameters as \cref{fig:loss-function}, and we include a low-energy cutoff of \SI{10}{\meV} as a stand-in for a finite detector threshold. Both the overall modulation amplitude and the dependence on the DM mass should only be taken for illustration and not as representative values, since they depend strongly on the anisotropy parameters; see \cref{fig:fixed DM mass,fig:fixed anisotropy} below.
    }
    \label{fig:full Modulation}
\end{figure*}

Material anisotropy can lead to a significant modulation in the measured event rate depending on the direction of the incident DM. In \cref{eq:event-rate}, we connected the time-dependent event rate $R(t)$ of DM-electron scattering directly to the material response. As an illustrative example, we now consider the daily modulation of the event rate for the anisotropic electron gas presented in \cref{sec:aeg}. In the following sections, we perform all integrals numerically using the \texttt{VegasFlow} package \cite{Carrazza:2020rdn,vegasflow_package}.

We adopt the conventions of \refcite{Coskuner:2019odd} to describe the laboratory frame on the Earth. The velocity of the laboratory frame relative to the isotropic halo frame is given as a function of time by
\begin{equation}
    \label{eq:v-lab}
    \bb v_{\mathrm{lab}}(t) =
    v_{\mathrm{E}}
    \begin{pmatrix}
        \sin\theta_e\sin\psi(t) \\
        \sin\theta_e\cos\theta_e(\cos\psi(t) - 1) \\
        \cos^2\theta_e + \sin^2\theta_e\cos\psi(t)
    \end{pmatrix}
    ,
\end{equation}
where $v_{\mathrm{E}}\simeq \SI{234}{\kilo\meter/\second}$ characterizes the motion of the Earth with respect to the DM halo; $\theta_e\approx 42^\circ$ is measured relative to the Milky Way frame; and $\psi(t) = 2\pi\times (t/\SI{24}{\hour})$. The convention employed here orients $\bb{v}_{\mathrm{lab}}$ entirely along the $\hat{\bb z}$ direction at $t=0$, changing to an orientation mostly along the $\hat{\bb y}$ direction at the 12 hour mark. For clarity, the direction of the velocity in \cref{eq:v-lab} is plotted over a 24 hour period in the left and center panels of \cref{fig:v-lab}.

\subsection{DM stream modulation}

As a check of our intuition, we first consider a DM distribution corresponding to a stream with fixed velocity $\bb v_s$ in the Milky Way frame,
\begin{equation}
\label{f halo example}
f_{\mathrm{halo}}(\bb v_\dm) = \delta^3(\bb v_\dm - \bb v_{s})
\,\,\,.
\end{equation}
In the laboratory frame, we obtain the DM velocity distribution as
\begin{equation}
    \label{eq:DM-lab-frame}
    f_{\mathrm{lab}}(\bb v_\dm,t) = f_{\mathrm{halo}}(\bb v_\dm + \bb v_{\mathrm{lab}}(t))\,\,\,,
\end{equation}
where the velocity of the lab frame $\bb v_{\mathrm{lab}}(t)$ is as in \cref{eq:v-lab}. The integral over the DM velocity for $g_0$ in \cref{eq:g0} is trivial for the distribution in \cref{eq:DM-lab-frame}, which directly provides the time-dependent event rate through \cref{eq:event-rate} as
\begin{equation}
    \label{event rate example}
    R(t) = \frac{1}{\rho_T}\frac{\rho_\dm}{m_\dm}
        \frac{\pi\bar\sigma}{\mu_{e\dm}^2}
        \int\frac{\du^3\bb q}{(2\pi)^3}
        \,
         \mathcal F(q)^2
            S\lp\bb q, \,\omega_{\bb q}(\bb v_s+\bb v_{\mathrm{lab}}(t)\rp
            .
\end{equation}

The right panel of \cref{fig:v-lab} shows the time-dependent event rate of \cref{event rate example} calculated with the electronic anisotropy and DM kinematics featured in \cref{fig:loss-function}. The DM stream velocity was chosen to be $\bb v_s=v_0\,\hat{\bb z}$, such that $\bb v_s\parallel \bb v_{\mathrm{lab}}(t)$ at $t=0$ and $v_0 = \SI{220}{\kilo\meter/\second}$ is the DM velocity dispersion.  We considered both the case of a massless mediator ($\mathcal{F}(q) \propto 1/q^2$) and the limit of a heavy mediator ($\mathcal F(q) = 1$) for the DM-electron interaction in \cref{eq:potential-definition}.  In both cases, the scattering rate exhibits the expected daily modulation due to the changing orientation of the Earth-based laboratory frame relative to the DM halo.

\subsection{Modulation in the Standard Halo Model}

A more realistic model of the DM velocity distribution is provided by the Standard Halo Model. To integrate over the distribution, we follow \refcite{Trickle:2019nya} to write the time-dependent event rate as in \cref{eq:event-rate}, with explicit expressions for $f_{\mathrm{halo}}(\mathbf{v}_\chi)$ and $g_0$ given in \cref{sec:SHM}.

In \cref{fig:full Modulation}, we show the daily modulation in our anisotropic electron gas target for the same material parameters as \cref{fig:loss-function}, but for several values of the DM mass in units of the electronic DOS mass $M$ through the ratio $m_\chi/M$. We retain the same low-energy cutoff $\omega_{\mathrm{min}}=\SI{10}{\meV}$ in performing the integration of \cref{eq:event-rate} to simulate a finite detector threshold. For both the light and heavy mediator, the daily modulation follows the same trend as the DM stream example in \cref{fig:v-lab}: the event rate is maximal at $t=0$ when the incident DM is mostly oriented along the $\hat{\bb z}$-axis and minimal at the 12 hour mark when DM is biased toward the $\hat{\bb y}$-axis. However, the peak-to-trough modulation amplitudes vary considerably. Though our fiducial anisotropy parameter of $m_z/m_{xy} = 20$ does not correspond to a particular physical system, and thus our modulation amplitudes should be taken for illustrative purposes only, they are typical of values for electronic excitations in anisotropic targets~\cite{Blanco:2021hlm,Kahn2022}.\footnote{One conspicuous exception is a Dirac material, where the linear dispersion makes scattering kinematically forbidden for $v_\chi < v_\fermi$ rather than simply suppressed~\cite{Hochberg:2017wce}, leading to a much larger modulation~\cite{Coskuner:2019odd}, though it is important to note that these analyses ignored any plasmon contributions.} 

\begin{figure}
    \centering
    \includegraphics[width=\columnwidth]{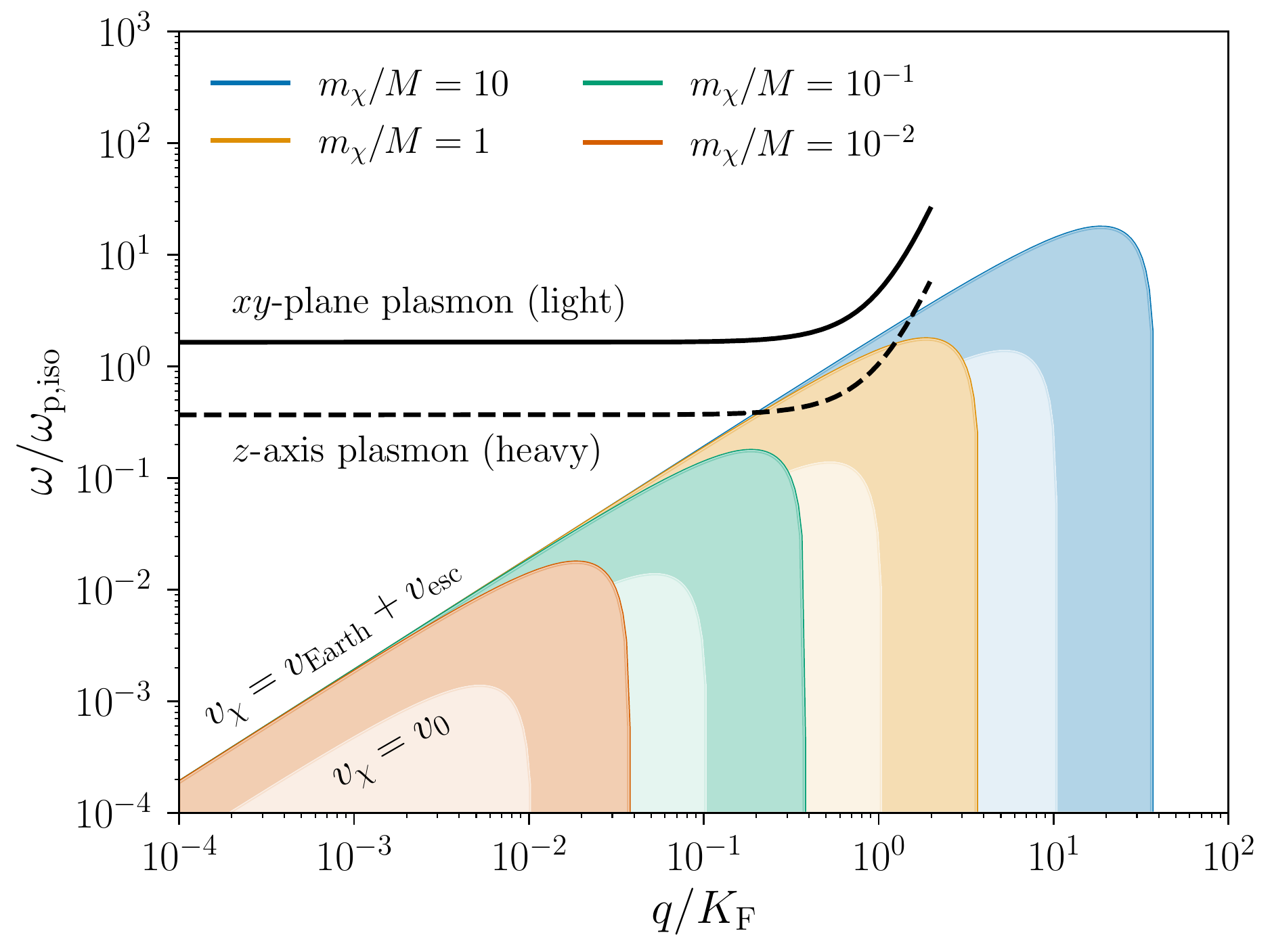}
    \caption{
    Kinematically-allowed regions for DM scattering, bounded by $\omega_{\bb q}(\bb v_\chi)$ of \cref{eq:dm-energy-loss}, for various DM masses compared to the DOS mass. The dark shaded regions take the DM velocity to be the maximum velocity in the lab frame, $v_\chi = v_{\mathrm{Earth}} + v_{\mathrm{esc}}$, while the light shaded regions show a typical DM velocity from the bulk of the distribution, $v_\chi = v_0$.  The black solid lines illustrate the  plasmon dispersions from \cref{fig:loss-function}, taking the same material and anisotropy parameters. Note that even though the higher-energy plasmon branch is \emph{energetically} accessible ($\Omega_{xy} < \frac{1}{2}m_\chi v_\chi^2$) for the heaviest DM masses, it is \emph{kinematically} inaccessible because it disperses before crossing into the DM parabola.
    }
    \label{fig:DM kinematics}
\end{figure}

To better understand the modulation curves shown in \cref{fig:full Modulation}, we show in \cref{fig:DM kinematics} the kinematically-allowed DM parabola with the plasmon dispersion shown for reference. For a massless mediator, the rate is peaked at small $q$, where the plasmons are the dominant excitation in the RPA response function. When the DM kinetic energy is small or comparable to the electronic scales (i.e., $E_\fermi$ and the isotropic plasma frequency $\omega_{\plas,{\iso}}$), the daily modulation in \cref{fig:full Modulation} is largely a probe of the anisotropic low-energy tail of the plasmon. When the lower branch of the plasmon is kinematically accessible ($m_\chi/M = 1$), the plasmon width is large enough that essentially any direction of $\mathbf{q}$ probes the plasmon part of the DM loss function, and the modulation amplitude decreases considerably. Note that independent of the value of $m_\chi$, our choice of parameters results in the upper branch of the plasmon along the $\hat{\bb x}$-axis \emph{always} being kinematically inaccessible; for larger DM masses, the apex of the DM energy-loss parabola shifts to the right, but the left boundary of the curve never crosses the plasmon before it damps out at large $q$. This is essentially because the effective Fermi velocity in the $xy$-plane is larger than $v_{\mathrm{max}}$, while the Fermi velocity in the $z$-direction is smaller than $v_{\mathrm{max}}$. For heavy mediators, the DM loss function is dominated by particle-hole excitations, which inherit their anisotropy directly from the dispersion relation of \cref{eq:aeg-dispersion} and are thus present even for large DM masses, leading to a large modulation amplitude for the same DM mass compared to a light mediator.

\begin{figure*}
    \centering
    \includegraphics[width=\textwidth]{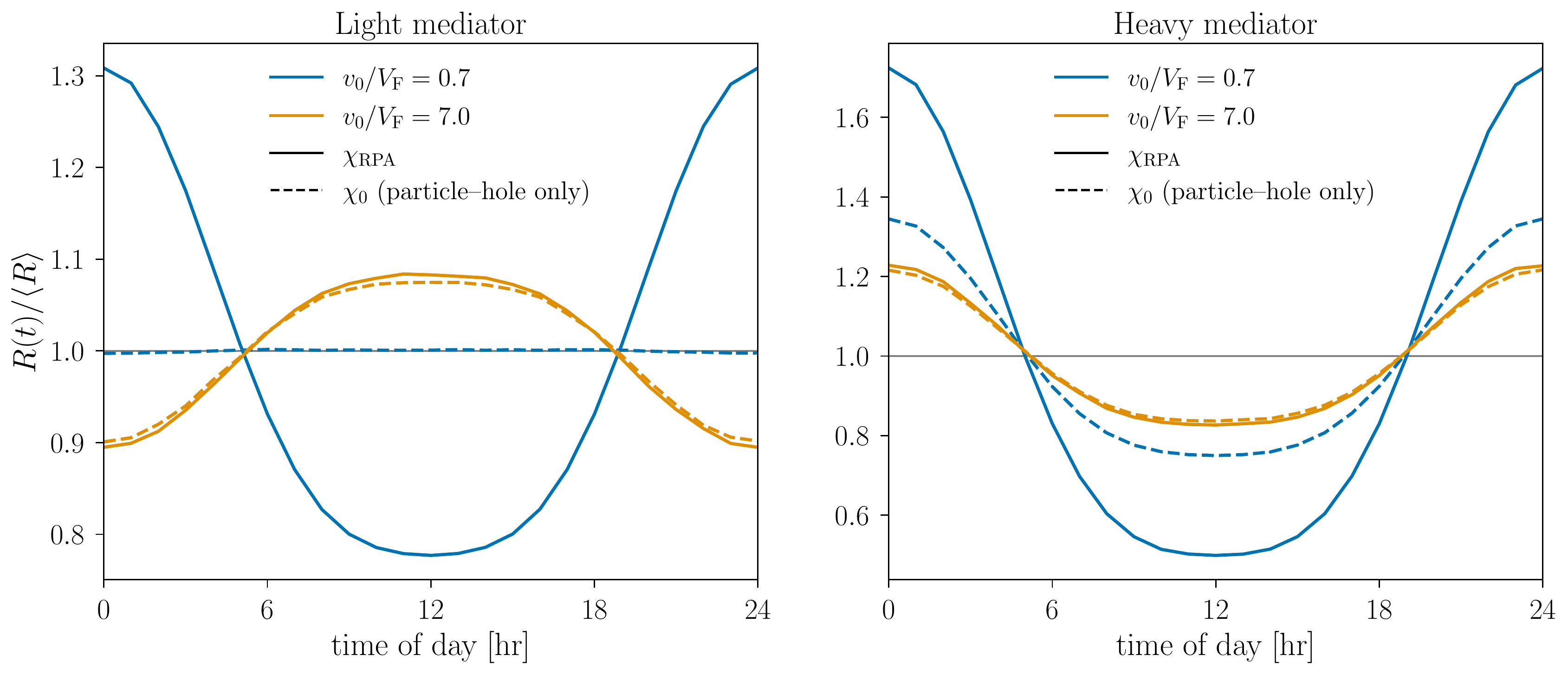}
    \caption{
    Daily modulation curves for $m_z/m_{xy} = 20$ but two different values of the isotropic Fermi velocity, for both a light mediator \textbf{(left)} and heavy mediator  \textbf{(right)}. Solid lines use $\chi_{\mathrm{RPA}}(\bb q,\omega)$ as in \cref{fig:fixed DM mass}, while dashed lines correspond to a dynamic structure factor $S(\bb q,\omega)$ determined solely by the \emph{bare} response $\chi_0(\bb q,\omega)$ in \cref{eq:anisotropic-to-isotropic}.  The use of $\chi_0(\bb q,\omega)$, rather than $\chi_{\mathrm{RPA}}(\bb q,\omega)$, restricts losses to particle-hole excitations and neglects the plasmon. The effect is largest for a light mediator with $v_0 < V_\fermi$, when scattering is dominated by the low-energy tail of the lower plasmon branch.
    }
    \label{fig:heavy bare modulation}
\end{figure*}

The effect of particle-hole excitations on the daily modulation can be isolated from the plasmon contribution by considering losses associated with the \emph{bare} density response function, $\chi_0(\bb q,\omega)$ of \cref{eq:chi0-definition}.  In \cref{fig:heavy bare modulation}, we show daily modulation curves computed from \cref{eq:event-rate} for both light and heavy mediators, for two different values of $V_\fermi$, comparing the rate computed using a dynamic structure factor $S(\bb q,\omega)$ determined through \cref{eq:structure-factor-density-response} by $\chi_0(\bb q,\omega)$ to the full RPA density response function $\chi_{\mathrm{RPA}}(\bb q,\omega)$.  For light mediators, when the lower branch of the plasmon is kinematically accessible ($v_0/V_\fermi = 0.7$), the rate is dominated by the plasmon and including only $\chi_0$ results in effectively zero modulation. For heavy mediators, the effect is much smaller, at the percent level for DM kinetic energies well above the Fermi energy ($v_0/V_\fermi = 7.0$). In fact, in this regime the particle-hole excitations dominate even the light-mediator loss function, though curiously the modulation changes phase, with a maximum at 12 hr.

\subsection{Relation between modulation amplitude and material parameters}

In order to clarify the parametric dependence of the daily modulation illustrated in \cref{fig:full Modulation}, we consider the effect of varying the material and anisotropy parameters on the daily modulation curves through a simple measure of variation in the time-dependent event rate: the peak-to-trough modulation
\begin{equation}
\label{peak to trough}
    \Delta_R =\frac{R(t=0)-R(t=12\text{ hours})}{\langle R\rangle}
    \,\,\,,
\end{equation}
where $\langle R\rangle$ is the time-averaged event rate over a 24 hour period. 

\begin{figure*}[t!]
    \centering
    \includegraphics[width=\textwidth]{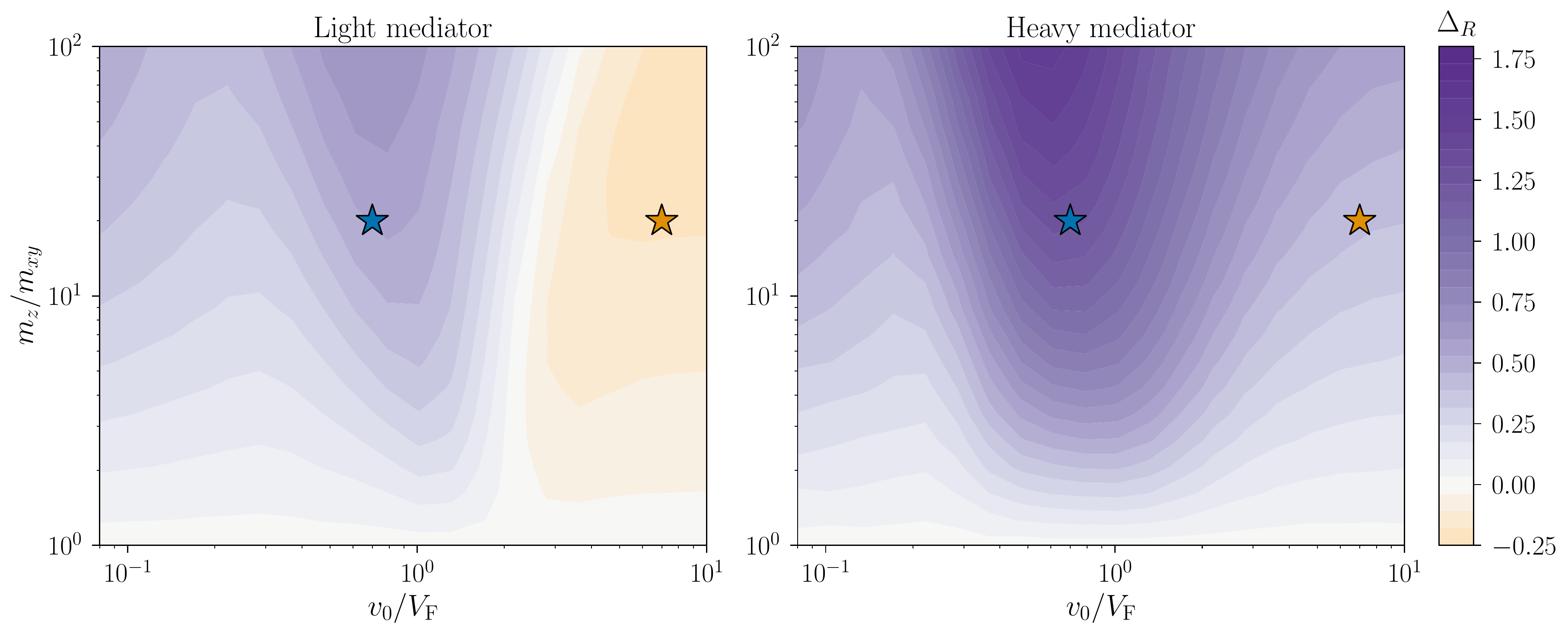}
    \caption{
    Density plots of the peak-to-trough modulation $\Delta_R$ of \cref{peak to trough} as a function of the uniaxial electronic mass anisotropy $m_z/m_{xy}$ ($m_{xy}=m_x=m_y$) and the velocity ratio $v_0/V_\fermi$, where the DM mass is set equal to the target DOS mass $m_\chi=M=m_e$.  The SHM event rate $R(t)$ was calculated using $\chi_{\mathrm{RPA}}(\bb q,\omega)$ of \cref{eq:model-RPA}, the values of $m_z/m_{xy}$ and $v_0/V_\fermi$ shown, and the plasmon width according to $\Gamma_\plas=0.1\,E_\fermi$.  Stars denote special points at $m_z/m_{xy}=20$ examined in \cref{fig:heavy bare modulation}.  When using a light mediator \textbf{(left)}, the small $q$ regime is enhanced, which coincides with the regime of plasmon propagation in \cref{fig:loss-function}.  When using a heavy mediator \textbf{(right)}, the plasmon contribution is suppressed at small $q$ and particle-hole excitation is enhanced at large $q$. The stars mark the parameter values chosen to generate the daily modulation curves in \cref{fig:heavy bare modulation}.
    }
    \label{fig:fixed DM mass}
\end{figure*}

In \cref{fig:fixed DM mass}, we plot the modulation amplitude at a fixed DM mass, $m_\chi=M = m_e$, varying the isotropic Fermi velocity $V_\fermi$ from $0.1 v_0$ to $10 v_0$ and anisotropy parameter $m_z/m_{xy}$ from 1 to 100 while preserving the relationship $\omega_{\plas,\iso} = E_\fermi$ by changing $\epsilon_\infty$ to compensate. In particular, the range of Fermi velocities corresponds to target Fermi energies $E_\fermi = \frac{1}{2}M V_\fermi^2$ spanning \SI{1.4}{\meV} to \SI{14}{\eV}, which sets the typical scale of both plasmons and particle-hole excitations. In the regime where the typical scale of plasmon excitation is kinematically inaccessible, $v_0/V_\fermi \lesssim 1$, the peak-to-trough ratio can be understood in the same way as the modulation of \cref{fig:full Modulation}: momentum transfer along the $\hat{\bb z}$ direction corresponds to smaller, more favorable energy scales for particle-hole or plasmon excitation, which results in modulation amplitudes which grow with the anisotropy parameter $m_z/m_x$. For a massless mediator (left panel), there is a rather sharp dividing line at $v_0/V_\fermi = 1$; for smaller Fermi velocities compared to the DM velocity, the plasmon peaks become kinematically accessible as illustrated in \cref{fig:DM kinematics} which increases the total rate but decreases the modulation amplitude. This results in a separatrix where for sufficiently small $V_\fermi$ the modulation $\Delta_R$ goes through a sign change. These effects are less pronounced for a massive mediator (right panel), where the preference for large momentum transfers and the dominance of particle-hole excitations results in a balancing act wherein the daily modulation increases with increasing $v_0/V_\fermi\lesssim1$, achieving a maximum at $v_0/V_\fermi \simeq 1$. The range of typical behavior for $m_z/m_{xy} = 20$ is illustrated in \cref{fig:heavy bare modulation} with the parameter points marked with stars.

\begin{figure*}[t!]
    \centering
    \includegraphics[width=\textwidth]{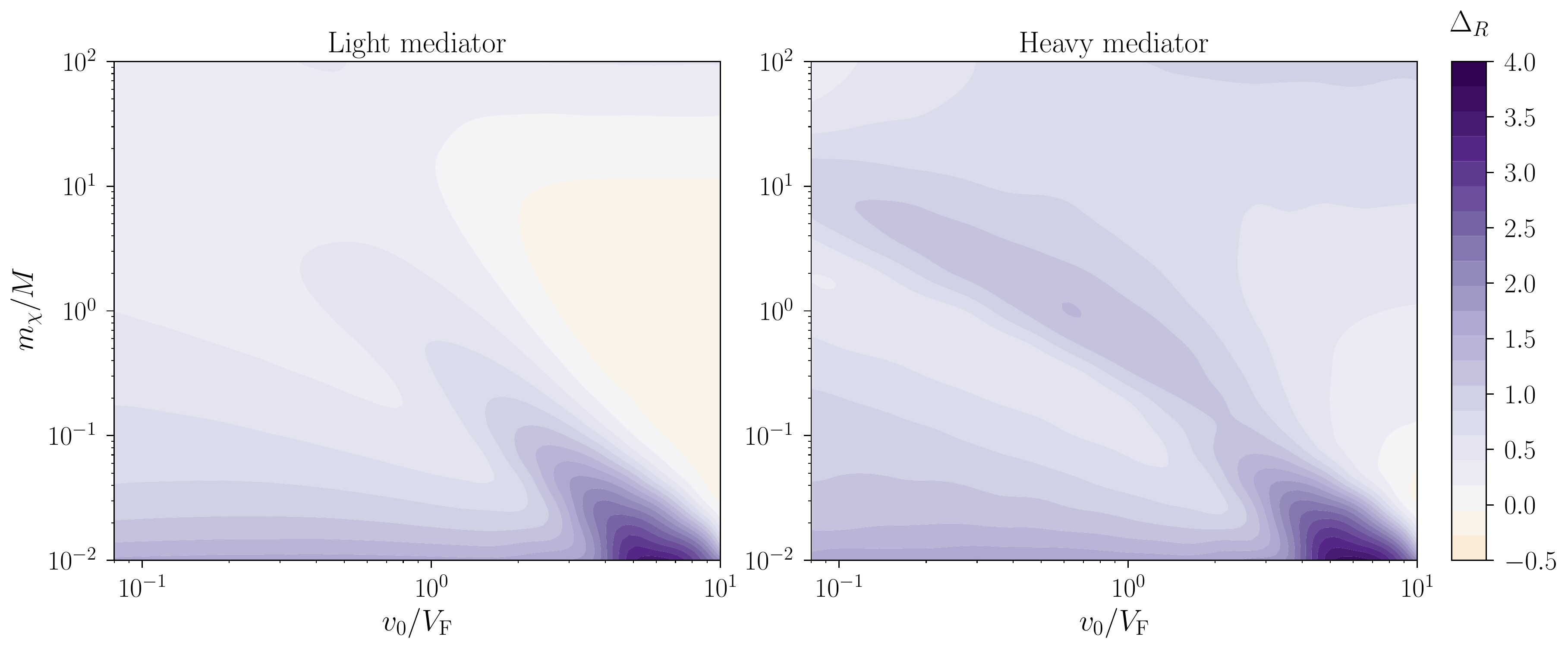}
    \caption{
    Same as \cref{fig:fixed DM mass} but fixing the electronic mass anisotropy to $m_z/m_{xy}=20$ and varying the DM mass ratio $m_\chi/M$ along the vertical axis. With the exception of an ``island'' at $m_\chi \simeq M$ and $v_0 \simeq V_\fermi$, the modulation amplitude typically decreases as the DM mass increases due to the absence of kinematic thresholds. The large modulation amplitude at large $v_0/V_\fermi$ and small $m_\chi/M$ is a threshold effect when DM does not have enough kinetic energy to create particle-hole excitations in the light effective mass directions, however this parameter space corresponds to atypically small Fermi energies.
    }
    \label{fig:fixed anisotropy}
\end{figure*}

In \cref{fig:fixed anisotropy}, we now fix the anisotropy parameter to the same fiducial value used previously, $m_z/m_x=20$ and vary the DM mass (normalized to the DOS mass, $m_\chi/M$) as well as $V_\fermi$. For a DOS mass $M$ on the order of the electron mass, $M\sim m_e\sim O$(\SI{}{\MeV}), our choice of scales corresponds to light DM masses from \SI{10}{\keV} to \SI{100}{\MeV}. We see that the modulation is appreciably smaller for heavy DM where there are fewer kinematic thresholds in either the particle-hole or plasmon spectra, but for the same DM mass, heavy-mediator models result in larger modulation amplitudes for the same DM mass and Fermi velocity due to the dominance of particle-hole excitations. The largest modulation occurs when $m_\chi$ is small and $v_0/V_\fermi$ is large, corresponding to particle-hole excitations with a very small Fermi energy and a kinematic threshold for excitations along the heavy effective mass directions. Because this is a threshold effect, the total rate is small. On the other hand, when the DM mass $m_\chi$ and velocity $v_0$ are well-matched to the analogous target scales $M$ and $V_\fermi$, order-1 modulation with large total rates are achievable, which highlights the importance of identifying suitable target materials which realize these scales.

\section{Extracting the density response from experiments}
\label{sec:calibration}

Our prediction for an observable daily modulation in the DM scattering rate relies on detailed knowledge of the target density response at kinematically-allowed values of $\bb q$ and $\omega$.  We emphasize that the desired response of a solid state target is, in principle, an experimental observable; our analysis of the anisotropic electron gas calculation considered in \cref{sec:aeg} merely provided a convenient toy model.  As discussed in \cref{sec:scattering}, DM-electron and electron-electron scattering are both governed by the \emph{same} density response function of the target material, which renders EELS experiments a direct means of determining the DM scattering rate for a given choice of the DM-electron interaction, $V(q)$ of \cref{eq:potential-definition}.  Despite this overlap, extracting the momentum-resolved density response of potential targets likely requires a great deal of laboratory ingenuity.  In this section, we will briefly discuss the feasibility of using experimental data within the daily modulation formalism of \cref{sec:daily-modulation}.

The ability to use EELS data in the calculation of the DM-electron scattering rate depends largely on the energy scales of interest.  At energies of $\mathcal{O}(\SI{10}{\eV})$, an orientation-resolved scan of the plasmon anisotropy at fixed momentum transfer has been performed with EELS in graphite \cite{Fink2002}, a layered compound consisting of weakly-bound carbon planes. Further in \refcite{Fink2002}, good agreement was obtained between computational models and the measured spectra. If several values of the magnitude $q$ were additionally scanned, then one could entirely map out the EELS loss function of \cref{eq:loss-response-dielectric} and, through \cref{eq:structure-factor-density-response}, the dynamic structure factor, $S(\bb q,\omega)$.  For light DM carrying kinetic energy on the meV scale, however, the finite energy resolution of EELS---typically described by the width of the zero-loss peak or quasi-elastic line---of $\mathcal{O}(10\textnormal{--}\SI{100}{\meV})$ \cite{Fink2002,Egerton2011,Fink2014} becomes non-negligible.  In practice, the zero-loss peak may remain the dominant feature in measured spectra significantly beyond the stated width.  This effect can be seen in the uncorrected spectra of \refcite{Fink2014} and provides the motivation for computational means of removing zero loss peaks out to $\sim$\SI{1}{\eV} in \refscite{Roest2021,Brokkelkamp2022} despite the width of some spectrometers set as narrow as 20--\SI{50}{\meV}.  

\begin{figure*}[t!]
    \centering
    \includegraphics[width=\textwidth]{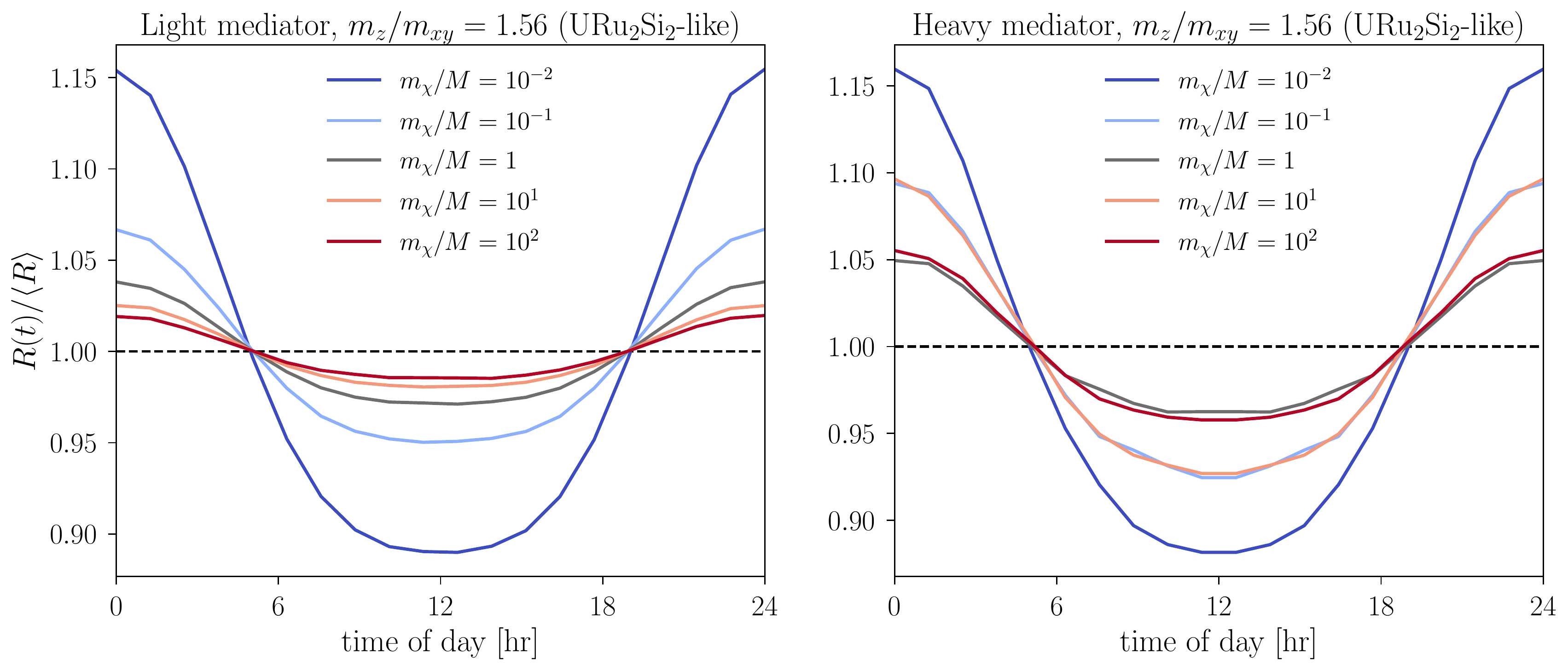}
    \caption{
    Daily modulation curves for DM scattering in an anisotropic electron gas, with anisotropy parameter $m_z/m_{xy}$ chosen to model the anisotropic $q \to 0$ loss function in the material \URuSi~\cite{vanderMarel2016_2}. In \refcite{Hochberg:2021pkt}, the low-energy part of the loss function was modeled as a plasmon with fixed frequency $\omega_\plas$ but anisotropic width; here, for illustration, we fix the width and imagine the anisotropic plasmon frequencies are inherited from an anisotropic effective mass. We take $E_\fermi = \omega_{\plas,\iso} = \SI{13.7}{\eV}$ and $M = m_e$, treating the plasmon as sourced by the full valence electron density rather than the heavy-fermion portion. Since $v_0 < V_\fermi$ and the anisotropy parameter is rather small, the modulation amplitude is only at the 20\% level even for small DM masses.}
    \label{fig:fast modulation}
\end{figure*}

Nevertheless, a measurement-driven approach could be maintained for the low-energy tail of an anisotropic plasmon.   Since the plasmon energy is typically on the order of a few to tens of eV in normal metals, one could evade the resolution issue in this low-energy regime by utilizing EELS at higher energies to accurately extract the anisotropy in the (electronic) dynamic structure factor $S(\bb q,\omega)$ of \cref{eq:event-rate} due to plasma losses.  If, as in \refcite{Hochberg:2021pkt}, we can model the plasmon peaks as Lorentzian, then we might obtain reliable results from a linear extrapolation of the measured spectra down to zero energy loss.\footnote{This is typically done in practice when using the Kramers--Kronig relation to extract the dielectric function from an EELS measurement.} In \cref{fig:fast modulation}, we show that \SI{50}{\keV} DM with kinetic energies on the order of $\mathcal O\lp\frac12 m_\chi v_0^2\rp\approx\SI{14}{\meV}$ still scatters with appreciable daily modulation.  This is despite the small anisotropy chosen to match the $q \to 0$ loss function in \URuSi~\cite{vanderMarel2016_2}, as first considered in \refcite{Hochberg:2021pkt}, and when the bulk plasmon energy scale (set to $\omega_{\plas,\iso}\approx13.7$ eV) is kinematically inaccessible.  Of course, the total event rate of DM-electron scattering will be suppressed when the scattering kinematics are mismatched, which serves to reinforce the point that daily modulation and total rate are typically in tension.

As suggested in \refcite{Hochberg:2021pkt}, the total DM scattering rate in a laboratory experiment could be increased through use of novel solid-state targets with energy scales (e.g., the plasma frequency and/or Fermi velocity in metals) more tuned to the DM kinematics.  For instance, heavy-fermion materials may feature plasmons at the meV scale, albeit with a lower spectral weight than expected of the higher-energy plasmons in normal metals. At the meV scale, however, the aforementioned energy resolution in EELS will likely prevent direct extraction of the loss function.  Additionally, the ionic contributions that have thus far been ignored in our analysis typically contribute to the charge density response (e.g., as measured in EELS) across the 1--\SI{100}{\meV} range.

One may then be tempted to look toward \emph{High-Resolution} EELS---i.e., HREELS \cite{Ibach1977,Ibach1982}, or its momentum-resolved (MEELS) \cite{MEELS} implementation.  While HREELS boasts an impressive energy resolution at the meV scale \cite{Ibach1996}, the scattering process in HREELS involves electron \emph{reflection} off the material surface.  This is in contrast to bulk scattering deep within a sample, which is the mechanism considered within our analysis of the DM-electron scattering rate in \cref{eq:dm-scattering-rate}.  Instead, the reflection process in HREELS is modeled by vacuum scattering off the long-ranged dipole field of the material \cite{Mills1972,Mills1975,MEELS} and the inclusion of the surface plane restricts translation symmetry to only the planar directions.  Naturally, both EELS and HREELS probe the density response function of a necessarily-finite sample with boundaries; HREELS, however, is dominated by processes in which the surface plane's reflective contribution is fundamental.  Nevertheless, the HREELS cross section is a probe of bulk physics: the strongly anisotropic \emph{bulk} plasmon of layered materials is predicted to be the main contribution to the long wavelength HREELS spectrum \cite{Boyd2022}.  While not model-independent (since bulk scattering requires extrapolation) and reliant on cleavable surface planes along the momentum directions of interest, HREELS can, in principle, provide data-driven predictions of DM scattering rates.

As long as one is willing to sacrifice momentum-dependent information for energy resolution, optical probes provide another avenue of investigation.  Optical methods---in particular, reflectance or ellipsometry~\cite{Wooten1972,Dressel2002}---provide the dielectric function, $\e(\omega)$, either via Kramers--Kronig relations (reflectance) or by model substitution of the measured quantities (ellipsometry), which permits calculation of the EELS loss function in \cref{eq:loss-response-dielectric} in the $q\to0$ limit.  Though typically limited to energies $\omega \gtrsim 10\textnormal{--}\SI{20}{\meV}$ and/or requiring Kramers--Kronig extrapolations, optical methods can provide meV-scale resolution. Optical probes do not offer the ability to predict losses at finite momentum transfer; however, the anisotropy of long-wavelength losses can still be determined through the reflectance of polarized light at normal incidence, as done in \refcite{vanderMarel2016_2} for the novel heavy-fermion compound \URuSi (in addition to more involved ellipsometric analyses).  As a result, optical experiments provide a high-resolution, long wavelength check on more involved experiments or modeling efforts of both the scale of losses and the material anisotropy.

\section{Conclusions}
\label{sec:conclusions}
In this paper, we have laid out a formalism which permits the derivation of the time-dependent scattering rate for spin-independent DM-electron interactions in an arbitrary target material based only on in-principle measurable quantities. By focusing exclusively on response functions---in particular the anisotropic density response function---we have established an important connection between quantities which can be measured in a laboratory setting via calibration experiments and the predicted daily modulation of the DM rate, thus removing a large source of systematic uncertainty from current and planned experiments. As an example, we have used the toy model of an anisotropic electron gas, where a semi-analytic treatment of the density response is tractable;  this model captures important collective effects (in particular, the anisotropic plasmon) which should reflect the behavior of real materials in an appropriate energy and momentum regime.

Extracting the density response function from measurements may be a highly non-trivial task due to the energy scales of interest and the need for excellent momentum resolution, and may not be entirely free of modeling uncertainties due to the need to account for surface effects in scattering experiments. Nonetheless, there are realistic experimental probes such as MEELS and HREELS that can be applied to materials of interest. We emphasize that even in the case of well-studied materials such as silicon, the density response has not been probed at the required accuracy in the kinematic regime relevant for DM scattering, and we encourage the application of such probes to the materials currently being used as light DM detectors to complement existing theoretical treatments of the DM rate. While our work here has focused on spin-independent scattering, a similar approach can be applied to other forms of DM-SM interactions (such as those categorized in Ref.~\cite{Trickle:2020oki}), which we leave for future work. We hope that our work continues to facilitate a rapid exchange of ideas and data between the high-energy and condensed matter communities in the development of new DM detectors.

\begin{acknowledgments}
    We thank Peter Abbamonte, Daniele Alves, Michael Graesser, Christopher Lane, Elizabeth Peterson, Sam Watkins, and Jianxin Zhu for helpful conversations. This work was partially supported by the Laboratory Directed Research and Development program of Los Alamos National Laboratory under project number 20220135DR. The work of Y.K.\ is supported in part by DOE grant DE-SC0015655. The work of Y.H.\ is supported by the Israel Science Foundation (grant No.\ 1818/22), by the Binational Science Foundation (grant No.\ 2018140),  by the Azrieli Foundation and by an ERC STG grant (grant No.\ 101040019). N.K. is supported in part by the US. Department of Energy Early Career Research Program (ECRP) under FWP 100872. The work of B.V.L.\ is supported by DOE grant No.\ DE-SC0010107, by the Josephine de Karman Fellowship Trust, and by the MIT Pappalardo Fellowship. This project has received funding from the European Research Council (ERC) under the European Union’s Horizon Europe research and innovation programme (grant agreement No.\ 101040019).  Views and opinions expressed are however those of the author(s) only and do not necessarily reflect those of the European Union. The European Union cannot be held responsible for them.
\end{acknowledgments}

\appendix

\section{The SHM time-dependent velocity distribution in the lab frame}
\label{sec:SHM}

Given a DM velocity distribution in the Galactic frame $f_{\mathrm{halo}}(\bb v_\chi)$, the time-dependent lab distribution of DM velocities is
\begin{equation}
    \label{f lab}
    f_{\mathrm{lab}}(\bb v_\dm,t) = f_{\mathrm{halo}}(\bb v_\dm + \bb v_{\mathrm{lab}}(t))
    \,\,\,,
\end{equation}
where $\bb v_{\mathrm{lab}}(t)$ is the velocity of the lab frame as given in \cref{eq:v-lab}.  Integrating the velocity-resolved scattering rate $\Gamma$ of \cref{eq:dm-scattering-rate} over the lab-frame DM velocity distribution yields the time-dependent event rate per unit detector mass,
\begin{equation}
\label{event rate velocity}
    R(t) = \frac{1}{\rho_T}\frac{\rho_\dm}{m_\dm}
        \int\du^3\bb v_\dm \,f_{\mathrm{lab}}(\bb v_\dm,t) \Gamma(\bb v_\dm)
    \,\,\,,
\end{equation}
where $\rho_T$ is the scattering target density. Following \refcite{Trickle:2019nya}, the energy-conserving delta function in the scattering rate of \cref{eq:dm-scattering-rate} can be used to re-write \cref{event rate velocity} as
\begin{equation}
    R(t) = \frac{1}{\rho_T}\frac{\rho_\dm}{m_\dm}
        \frac{\pi\bar\sigma}{\mu_{e\dm}^2}
        \int\frac{\du^3\bb q\dd\omega}{(2\pi)^3}g_0(\bb q, \omega,t)
        \mathcal F(q)^2
            S(\bb q, \omega)
    \,\,\,,
\end{equation}
where we have introduced the anisotropic halo integral
\begin{equation}
\label{g0 def}
    g_0(\bb q, \omega,t) = \int\du^3\bb v_\dm\, f_{\mathrm{lab}}(\bb v_\dm,t) \,\delta\lp\omega - \omega_{\bb q}(\bb v_\chi)\rp
    \,\,\,.
\end{equation}

For illustration in this paper, we use the Standard Halo Model (SHM) ansatz of a truncated Maxwellian,
\begin{equation}
\label{f halo}
    f_{\mathrm{halo}}(\bb v_\dm) = N_0^{-1}\exp(-\bb v_\dm^2/v_0^2)\Theta(v^2_{\mathrm{esc}} - |\bb v_\dm|^2)
    \,\,\,,
\end{equation}
where we take the dispersion as $v_0\simeq 220$ km/s, the Galactic escape velocity is roughly $v_{\mathrm{esc}}\simeq544$ km/s, and the normalization is given by
\begin{equation}
    N_0=\pi^{3/2}v_0^3\lb
    \text{erf}\lp\frac{v_{\mathrm{esc}}}{v_0}\rp-\frac{2}{\sqrt\pi}\frac{v_{\mathrm{esc}}}{v_0} \exp\lp -\frac{v_{\mathrm{esc}}}{v_0}\rp^2
    \rb
    .
\end{equation}
The SHM ansatz gives a closed-form expression for the anisotropic halo integral~\cite{Griffin:2018bjn},
\begin{equation}
\label{g0}
    g_0(\bb q, \omega,t) = \frac{\pi v_0^2}{qN_0}\biggl[
    e^{-v_-^2(\bb q,t)/v_0^2}
    -
        e^{-v_{\mathrm{esc}}^2/v_0^2}
    \biggr]
    ,
\end{equation}
where
\begin{equation}
    v_-(\bb q,t)=\min\left\{
                v_{\mathrm{esc}}\,,
                \textstyle\frac \omega q + \frac{q}{2m_\dm} +
                \hat{\bb q}\cdot\bb v_{\mathrm{lab}}(t)
            \right\}
            \,\,\,.
\end{equation}
With the SHM halo integral in \cref{g0}, the time-dependent event rate for DM-electron scattering off anisotropic solid state targets can be obtained through $R(t)$ in \cref{eq:event-rate} provided we know the (electronic) dynamic structure factor of the solid state target, $S(\bb q,\omega)$, over the kinematically-allowed region of $\bb q$ and $\omega$.

\section{The dielectric function, electron scattering, and density response of materials}
\label{sec:material-response}

The relation between material response and DM scattering in solid state targets has developed rapidly in recent years; see \refcite{Kahn2022} and references therein for a review. However, there are some technical and formal aspects that may be more familiar to those with a background in solid-state physics rather than particle physics.  In this Appendix, we provide further details on the basic framework assumed in the main text and attempt to clarify nomenclature when possible. As in \cref{sec:scattering}, no original material is presented here, and the emphasis is on translation between the solid-state and particle physics languages. To maintain continuity with the main text, we adopt particle physics conventions, particularly through the system of units where $\hbar=c=\e_0=1$.  In \cref{app: dielectric}, we review the definition of optical constants in the framework of classical electromagnetism via Maxwell's equations in media. In \cref{app: scattering}, we review the framework of density-density scattering off the electrons within a solid-state target and explicitly connect the many-body density response function to the matrix elements that appear in Fermi's Golden Rule.

\subsection{The dielectric theory of materials}
\label{app: dielectric}

In the framework of classical electromagnetism, adopted by optics texts such as \cite{Wooten1972,Dressel2002}, the behavior of a material is characterized through constitutive equations that relate the charges, currents, and fields exterior to the material to those induced within it. In particular, we focus on a linear dielectric medium subject to an external field $\bb E$, which screens the field to produce an electric displacement.
\begin{equation}
\label{dielectric def}
\bb D=\tens\e  \bb E
\,\,\,.
\end{equation}
The associated current response is given by
\begin{equation}
\label{conductivity def}
\bb J_{\mathrm{ind}} = \tens\sigma  \bb E
\,\,\,.
\end{equation}
The linearized relations in \cref{dielectric def,conductivity def} define the optical constants $\tens\e $, the dielectric constant, and $\tens\sigma $, the optical conductivity, where our notation reflects the fact that both are generically tensor quantities. In \cref{dielectric def,conductivity def}: $\bb E$ is the (total) electric field, $\bb D$ is the part of the electric field sourced only by external charges not bound to the material system, and $\bb J_{ind}$ is the current density induced within the material by the (total) electric field.  Of course, $\tens\e $ and $\tens\sigma $ are not truly constants; when characterizing a spacetime translation-invariant system, these optical constants depend on the frequency and wavevector of light (or whatever electromagnetic field is present) within the material medium through the Fourier-transformed relations:
\begin{align}
    \label{app: optical dielectric function}
    &\bb D(\bb q,\omega) = \tens\e (\bb q,\omega) \bb E(\bb q,\omega)\,\,\,,
    \\
    \label{app: optical conductivity}
    &\bb J_{\mathrm{ind}}(\bb q,\omega) = \tens\sigma (\bb q,\omega) \bb E(\bb q,\omega)
    \,\,\,.
\end{align}

The continuity equation
\begin{equation}
    \label{continuity equation}
    \frac{\partial}{\partial t}\rho + \bb \nabla\cdot \bb J = 0\,\,\,,
\end{equation}
where $\rho$ is the local (total) charge density and $\bb J$ is the (total) current density, yields a relation between the optical constants, $\tens \e $ and $\tens \sigma $.  As we are separating external charges (those not bound to the material system) from the charges and currents induced within the material, we can restrict the continuity equation of \cref{continuity equation} to the material charges:
\begin{equation}
    \label{induced continuity equation}
    \frac{\partial}{\partial t}\rho_{\mathrm{ind}} + \bb \nabla\cdot \bb J_{\mathrm{ind}} = 0\,\,\,,
\end{equation}
where $\rho_{\mathrm{ind}}$ is the (local) deviation of the material charge density from its equilibrium value.  The Fourier-transformed Gauss laws governing the electric displacement,
\begin{equation}
    \label{Gauss law D}
    i\bb q\cdot \bb D(\bb q,\omega)
    =
    \rho_{\mathrm{ext}}(\bb q,\omega)\,\,\,,
\end{equation}
and the (total) electric field,
\begin{equation}
    \label{Gauss law E}
    i\bb q\cdot \bb E(\bb q,\omega) = \rho(\bb q,\omega)\,\,\,,
\end{equation}
permit us to write the induced charge density, $\rho_{\mathrm{ind}}$, as
\begin{align}
    \rho_{\mathrm{ind}}(\bb q,\omega) &= \rho(\bb q,\omega)-\rho_{\mathrm{ext}}(\bb q,\omega)
    \\
    &= i\bb q\cdot\lb
    \bb E(\bb q,\omega)-\bb D(\bb q,\omega)
    \rb
    \\ \label{rho ind}
    &=
    i\bb q\cdot\lb
    1-\tens\e (\bb q,\omega)
    \rb\cdot \bb E(\bb q,\omega)
    \,\,\,.
\end{align}
If we use \cref{app: optical conductivity} to write the induced current in terms of $\tens\sigma (\bb q,\omega)$, then \cref{rho ind} allows us to write the continuity equation, \cref{induced continuity equation}, as
\begin{equation}
    \label{longitudinal continuity}
    \bb q\cdot\lb
    \omega\lp1-\tens\e (\bb q,\omega)\rp
    +i\tens\sigma (\bb q,\omega)\rb\cdot \bb E(\bb q,\omega) = 0
    \,\,\,.
\end{equation}

In the non-retarded limit of electromagnetism, where the Coulomb interaction is effectively instantaneous and magnetic back-reaction is negligible, the electric field is determined by a scalar potential $\phi$ via
\begin{equation}
    \label{app: electric potential def}
    \bb E(\bb q,\omega) = \lp-i\bb q\rp\phi(\bb q,\omega)\,\,\ .
\end{equation}
This description is appropriate for the scattering regime of the main text, where $q \gg \omega$. If we write the continuity equation, \cref{longitudinal continuity}, in terms of the potential of \cref{app: electric potential def}, we find that the \emph{longitudinal} components of $\tens \e $ and $\tens \sigma $ are related by
\begin{equation}
\label{longitudinal optical relation}
    \bb{\hat q}\cdot\tens\e (\bb q,\omega)\cdot \bb{\hat q}
    =
    1+\lp\frac{i}{\omega}\rp \bb{\hat q}\cdot\tens\sigma (\bb q,\omega)\cdot \bb{\hat q}
    \,\,\,,
\end{equation}
where $\bb{\hat q}:=\bb q/q$.  Since the non-retarded limit admits a description in terms of potentials, it's natural to consider the longitudinal dielectric function,
\begin{equation}
\label{longitudinal eps relation}
    \e_\lon(\bb q,\omega)=\bb{\hat q} \cdot \tens\e (\bb q,\omega)\cdot \bb{\hat q}\,\,\,,
\end{equation}
which is a \emph{scalar} relating the external potential $\phi_{ext}$,
\begin{equation}
    \label{app: external potential def}
    \bb D(\bb q,\omega) = \lp-i\bb q\rp\phi_{\mathrm{ext}}(\bb q,\omega)\,\,\,,
\end{equation}
to the total potential, $\phi$, via
\begin{equation}
    \label{longitudinal eps def}
    \phi_{\mathrm{ext}}(\bb q,\omega) = \e_\lon(\bb q,\omega)\phi(\bb q,\omega)
    \,\,\,.
\end{equation}
While the relation in \cref{longitudinal eps relation} connects the longitudinal dielectric function, $\e_\lon$, to the previous discussion of optical constants (e.g., one could extract $\e_\lon$ from the optical conductivity upon additional use of \cref{longitudinal optical relation}), the relation that most naturally generalizes to many-body screening in the non-retarded limit is \cref{longitudinal eps def}.

Including retardation effects and transverse fields is slightly more subtle. For this, we consider the tensor structure of the dielectric function $\tens\e$, as well as the Maxwell equations:
\begin{align}
    \nabla \times \bb E = \dot {\bb B}\;,\qquad
    \nabla \times \bb B = \bb J + \dot {\bb D}\;.
\end{align}
Taking the curl of the first equation and inserting into the second gives, in Fourier space \cite{Rukhadze:1961},
\begin{align}
    (q^2 - \bb q \bb q - \omega^2 \tens\e)\bb E(\bb q,\omega) = i\omega \bb J_{\mathrm{ext}}(\bb q, \omega) \;.
\end{align}
Expressing $\bb E= - i \bb q \phi + i \omega \bb A$ in terms of potentials
then gives
\begin{align}
    \left[\tens\e - \frac{q^2-\tens{q q}}{\omega^2}\right](\bb q \phi - \omega \bb A) = \frac{\bb q}{q^2}\rho_{\mathrm{ext}}\;,
\end{align}
where it was assumed that there are no transverse currents (perpendicular to $\bb q$), and the longitudinal component of $\bb J$ was related to $\rho$ by continuity. By taking the transverse component of this equation, the vector potential $\bb A$ can be related to the gradient of $\phi$ through the dielectric tensor $\tens \e$ (in Coulomb gauge). This can then be re-inserted into the longitudinal component of the above equation. Some rearranging then gives
\begin{align}
    q^2\phi = \hat{\bb q} \cdot \left[\tens \e - \frac{q^2-\tens{q q}}{\omega^2}\right]^{-1}\cdot \hat{\bb q }\;\rho_{\mathrm{ext}}\;.
\end{align}
This yields an effective dielectric constant
\begin{align}
\label{eq:matinv}
    \epsilon_{\mathrm{eff}}^{-1} = \hat{\bb q} \cdot \left[\tens \e - \frac{q^2-\tens{q q}}{\omega^2}\right]^{-1}\cdot \hat{\bb q }
\end{align}
that includes retardation effects. Because of the matrix inverse, this is expression is actually dominated by the $\tens\e$ term rather than the $q^2/\omega^2$ term in the scattering limit ($q^2 \gg \omega^2$), and one can verify that the above expression indeed reduces to
\begin{align}
\label{app: inverse tensor dielectric}
\frac{1}{\epsilon_{\mathrm{eff}}} \simeq \frac{1}{\hat{\bb q}\cdot \tens \e \cdot \hat{\bb q}}
\end{align}
in this limit. Equivalent expressions to \cref{eq:matinv} involving the same matrix inverse have previously been derived for an electron moving through an anisotropic material~\cite{Tossati:1969}.

\subsection{Quantum mechanical treatment of electron scattering and the electron energy loss function}
\label{app: scattering}

In contrast to the classical description above, here we consider the scattering rate as determined by Fermi's Golden Rule for an electron scattering off a solid-state target.  The goal of our analysis is to connect with---and reproduce the basic components of---the literature on electron scattering as it pertains to Electron Energy Loss Spectroscopy (EELS).  The electron-electron coupling in EELS (excluding the ionic contribution to EELS in this analysis, as was done in the main text) is analogous to the DM-electron coupling in \cref{eq:interaction-hamiltonian} with the addition that the incoming and outgoing scattering states are (electron) plane waves. In other words, we are modeling electron scattering in the Born approximation, as is the standard assumption for DM scattering. Direct application of Fermi's Golden Rule to all possible excitations of the solid-state target, which we take to be in its ground state at zero temperature, results in the single-scattering rate
\begin{equation}
    \label{app: scattering rate}
    \Gamma_e(\bb q,\omega) = \frac{2\pi}{\Omega}\sum_f \Big| V_C(q) \bra{f} \hat n(-\bb q)\ket{0}\Big|^2\delta\lp\omega-\lb E_f-E_0\rb\rp
    \,,
\end{equation}
where $V_C(q)$ is the electron-electron Coulomb interaction defined in \cref{eq:coulomb-potential}, $\bb q$ is the momentum transfer to the target, $\omega$ is the energy loss between incoming and outgoing states of the scattered electron, and all other symbols are as defined for the dynamic (electronic) structure factor, $S(\bb q,\omega)$ of \cref{eq:structure-factor}.  Notably, this construction does not rely on any spatial symmetries of the (macroscopic) target.  As a consequence of our zero temperature analysis (as was done in the main text), the energy-conserving delta function in \cref{app: scattering rate} can only be satisfied for $\omega>0$, or for energy \emph{losses} of the scattered electron.  The particle physics convention for including the factor of $2\pi$ in $S(\bb q,\omega)$ provides, by comparison to \cref{eq:structure-factor}, the relation
\begin{equation}
    \label{app: Gamma S relation}
    \Gamma_e(\bb q,\omega) = \lb V_C(q)\rb^2 S(\bb q,\omega)
    \,\,\,
\end{equation}
for the electron scattering rate.

As emphasized in the (zero temperature) fluctuation--dissipation theorem in \cref{eq:structure-factor-density-response}, the many-body matrix elements governing electron scattering are precisely those that enter the density response function of the solid-state target.  In order to concretely demonstrate this relationship, we consider the Fourier transform of the density response function $\chi(\bb r, \bb r'; t)$,
\begin{equation}
\label{app: chi Fourier transform}
\chi(\bb q,t)=\frac{1}{\Omega}\int d^3 \bb r d^3 \bb r' e^{-i\bb q\cdot(\bb r-\bb r')}\chi(\bb r,\bb r';t)\,\,,
\end{equation}
where $\Omega$ is the material volume.  From the definition of $\chi(\bb r, \bb r';t)$ in \cref{eq:density-response-definition}, inserting the resolution of the identity (in the many-electron basis), $1=\sum_f\ket{f}\bra{f}$, gives
\begin{multline}
    \chi(\bb q,t)=\frac{-i\Theta(t)}{\Omega}\sum_f
    \Bigl[e^{-it(E_f-E_0)}\bra{0}\hat n(\bb q)\ket{f}\bra{f}\hat n(-\bb q)\ket{0}
    \\
    -e^{it(E_f-E_0)}\bra{0}\hat n(-\bb q)\ket{f}\bra{f}\hat n(\bb q)\ket{0}
    \Bigr]
    \,\,\,.
\end{multline}
Due to the adiabatic switching-on of the perturbing field within linear response theory, we obtain the frequency transform of $\chi(\bb q,t)$ as
\begin{multline}
    \label{fourier chi def}
    \chi(\bb q,\omega)=\frac{1}{\Omega}\sum_f\Big[
    \frac{\bra{0}\hat n(\bb q)\ket{f}\bra{f}\hat n(-\bb q)\ket{0}}{\omega+i\delta-\lp E_f-E_0\rp}
    \\
    -
    \frac{\bra{0}\hat n(-\bb q)\ket{f}\bra{f}\hat n(\bb q)\ket{0}}{\omega+i\delta+\lp E_{f}-E_0\rp}
    \Big]\,\,\,,
\end{multline}
where $\delta\to0^+$ isolates the causal response.  By use of the Dirac identity,
\begin{equation}
    \label{app: Dirac identity}
    \lim_{\e\to0^+}\im \frac{1}{x+i\e}=-\frac 1\pi\delta(x)
    \,\,\,,
\end{equation}
the $\delta\to0^+$ limit in \cref{fourier chi def} selects the imaginary part of $\chi(\bb q,\omega)$ as
\begin{multline}
    \label{imaginary chi}
    \im\chi(\bb q,\omega) =
    -\frac{\pi}{\Omega}\sum_f\Bigl[
    \Bigl|\bra{f}\hat n(-\bb q)\ket{0}\Bigr|^2
    \delta\lp\omega-\lb E_f-E_0\rb\rp
    \\
    -
    \Bigl|\bra{f}\hat n(\bb q)\ket{0}\Bigr|^2
    \delta\lp\omega+\lb E_f-E_0\rb\rp
    \Bigr].
\end{multline}
For $\omega>0$, the delta function of the second term in \cref{imaginary chi} identically vanishes and the first term gives exactly the same many-electron matrix elements that determine $S(\bb q,\omega)$ in \cref{eq:structure-factor}.  

The fluctuation--dissipation theorem in \cref{eq:structure-factor-density-response} allows us to write the scattering rate in \cref{app: scattering rate} in terms of the density response in \cref{imaginary chi} as
\begin{equation}
    \label{app: Gamma chi relation}
    \Gamma_e(\bb q,\omega)=\lb V_C(q)\rb^2 S(\bb q,\omega) = -2\lb V_C(q)\rb^2\im\chi(\bb q,\omega)
    \,\,\,.
\end{equation}
As such, the density response function of a solid-state target determines the material contribution to the EELS scattering rate.  At this stage, we have the tools to develop an anachronistic construction of the EELS loss function, \cref{eq:loss-response-dielectric}, wherein 
\begin{equation}
    \label{app: EELS loss function chi}
    \loss =-V_C(q)\im\chi(\bb q,\omega)
\end{equation}
provides a dimensionless quantity that, through \cref{app: Gamma chi relation}, is proportional to the spectrum measured by an EELS experiment. The proportionality factors are given by the translation between scattering rate and differential cross section, reviewed in \refscite{Sturm1993,Fink2014,Egerton2011}. As noted in \refscite{Hochberg:2021pkt,Knapen:2021run}, non-relativistic, light DM coupled to the target electron density scatters in the same way as \cref{app: scattering rate} if we replace $V_C(q)$ with the DM-electron potential, $V(q)$.  As a result, EELS and the EELS loss function---particularly through the construction in \cref{app: EELS loss function chi}---are a direct probe of the target contribution to DM-electron scattering that, for a given DM-electron interaction, requires no further modeling.

Historically, however, EELS was first understood through its relation to the \emph{optical} probes discussed in \cref{app: dielectric} above. Before the many-body machinery needed to approximate the density response function $\chi(\bb q,\omega)$ was established, the optical constants---$\tens\e (\bb q,\omega)$ of \cref{app: optical dielectric function} and  $\tens\sigma (\bb q,\omega)$ of \cref{app: optical conductivity}---were readily available for simple metals. Indeed, the ability to model EELS spectra via measured optical constants was recognized early on in \refcite{Frohlich1955}, where the optical constants of elemental silver were compared with scattering spectra at a time when the many-body formalism of electron scattering was still under development. As a result, EELS was interpreted through the electron energy loss function presented within a dielectric framework:
\begin{equation}
\label{loss function def}
\loss=-\im\e^{-1}_\lon(\bb q,\omega)
\,\,\,.
\end{equation}

By use of dielectric form of the loss function of \cref{loss function def}, the long-wavelength
 (i.e., the $q\to0$ limit with respect to material length scales) prescription of \eqref{app: inverse tensor dielectric} (see also \refcite{Tossati:1969})
 \begin{equation}
\label{app: optical to longitudinal}
\e^{-1}_\lon(\bb q,\omega)\overset{q\to0}\longleftrightarrow \frac{1}{\bb{\hat q}\cdot \tens\e (\omega)\cdot \bb{\hat q}}
\end{equation}
 was, and continues to be, capable of modeling EELS spectra through optical measurements. For example, the optical formula in \cref{app: optical to longitudinal} is essential to accurately model $\loss$ in \cref{loss function def} for the anisotropic, layered high-$T_C$ cuprate superconductor Bi-2212 \cite{vanderMarel2016} (compared directly to EELS data in \refcite{vanderMarel2015}) and the anisotropic, heavy-fermion compound \URuSi \cite{vanderMarel2016_2} discussed in the main text. Strictly speaking, however, optical methods probe the transverse response, rather than the longitudinal response encoded within \cref{loss function def,app: optical to longitudinal}.  In the $|\bb q|\to0$ limit, the longitudinal and transverse responses coincide in the absence of divergent magnetic response \cite{vanderMarel2016,Tossati:1969}.  Further, the dielectric loss relation \eqref{loss function def}, by use of \eqref{app: optical to longitudinal}, emphasizes that it is the \emph{inverse} dielectric function that is related to a causal response function (and, hence, scattering by the fluctuation--dissipation theorem) rather than the dielectric function itself (see, e.g., the discussion in \refcite{Dolgov1981}).

In summary, scattering of an external particle by the target electron density probes the (target, electronic) density response function, which is the physical content of the fluctuation--dissipation theorem in \cref{eq:structure-factor-density-response}.  We emphasize that this relationship does not require any particular spatial symmetry of the target, but does rely on a macroscopic sample whose boundaries can be neglected.  In EELS, the electron energy loss function, $\loss$ of \cref{eq:loss-response-dielectric}, captures the target contribution to Coulomb scattering.  In the context of DM scattering off the electron density within a material, EELS provides a direct measurement of the matrix elements that, for a given DM-electron interaction, determine the scattering rate unambiguously.  The dielectric framework, as previously discussed, readily connects density response (and, therefore, electron scattering) to long wavelength, optical measurements.

\bibliography{references}

\end{document}